\documentclass{mn2e}
\usepackage{natbib}
\usepackage{epsfig}
\usepackage{graphicx}
\begin{document}
\title[On Why Disks Generate Magnetic Towers and Collimate Jets]
  {On Why Disks Generate Magnetic Towers and Collimate Jets}
\author[D Lynden-Bell]
  {D Lynden-Bell\\
The Observatories of the Carnegie Institution,
813 Santa Barbara St., Pasadena, CA., U.S.A.\\
  The Institute of Astronomy, The Observatories, Madingley Road,
  Cambridge, CB3 0HA, U.K. (permanent address) \& Clare College.\\
Physics Dept., The Queens University, Belfast.}
\date{Accepted . Received }
\pagerange{\pageref{firstpage}--\pageref{lastpage}}
\pubyear{}
\label{firstpage}
\maketitle
\begin{abstract}
We show that accretion disks with magnetic fields in them ought to
make jets provided that their electrical conductivity prevents
slippage and there is an ambient pressure in their surroundings.

We study {\bf equilibria} of highly wound magnetic structures.
General Energy theorems demonstrate that they form tall magnetic
towers whose height grows with every turn at a velocity related to the
circular velocity in the accretion disk.  

The pinch effect amplifies the magnetic pressures toward the axis of
the towers whose stability is briefly considered.

We give solutions for all twist profiles $\Phi(P)=\Omega(P)t$ and for
any external pressure distribution $p(z)$.  The force--free currents
are given by ${\bf j} = \widetilde{\alpha}(P){\bf B}$ and we show that
the constant pressure case gives $\widetilde{\alpha} \propto
P^{-{\scriptscriptstyle{1\over 2}}}$ which leads to analytic solutions for
the fields.

\end{abstract}
\section{Introduction}\label{sec1}
To take some of the most dynamic objects in the universe in which
`apparently superluminal' motions have been seen and to work on such
problems using statics may seem an indication of a
seriously deranged scientist.  Nevertheless, the potential energies
involved in any problem can be studied statically and it is those
potential energies that drive the motion.  Thus, without studies of
the way the potential energies operate, the basic understanding of
{\bf why} the motions occur in the way they do may be lost.  As
\citet{ed26} said `the chief aim of the physicist in discussing a
theoretical problem is to obtain ``{\bf insight}'' -- to see which of
the numerous factors are particularly concerned in any effect and how
they work together to give it'. 
\begin{quote}
{\bf Even a perfect model of a phenomenon, that gives all the
  observables correctly, is not good science until it is analysed to
  show which aspects are essential for the phenomenon.}
\end{quote}
An unnecessarily detailed model, which reproduces the phenomenon, can
actually be a barrier to understanding.  The lack of real
understanding of what makes red--giants is a case in point!  It may
require the insight of an Eddington rather than the calculations of a
Chandrasekhar to simplify the model to the bare essentials.

Within an accretion disk any radial magnetic field will be sheared and
stretched by the differential rotation, so the resultant toroidal
magnetic field will grow until it is strong enough to arch up out of
the disk with the gas flowing back down.  

\citet{hoy60} were the first to describe this instability but the
conditions for its occurrence were more precisely calculated by
\citet{par65}.  In the resulting configuration the mass of gas in the
accretion disk continues to anchor and to twist the feet of the flux
tube.  We show here that if there is an ambient pressure in the
tenuous gas above the accretion disk then the flux tube will grow in
height with every twist of its feet but will not expand laterally.
Thus a tall tower of magnetic field is formed whose collimation or
aspect ratio $Z/R$ increases linearly with the number of twists.  

My thesis is that an important pre-requisite to understanding the
dynamics of the jets above accretion disks is a serious study of the
magneto-statics of the magnetic fields that they twist into their
coronae.  My first studies in this direction ended in total failure.
In 1979 I had a mechanism that would give a tall tower of magnetic
field that grew more collimated with every twist.  With much
enthusiasm I made a more detailed and exact calculation but to my
amazement and chagrin the field managed to expand to infinity and
disconnect itself after just over half a turn!  The hoped for
collimation that got better with every turn ended with only half a
turn when the degree of collimation was not a needle--like, one
degree, but a full 120 degrees! See Figure 1. 

\begin{figure}
\begin{center}
\includegraphics[scale=.65]{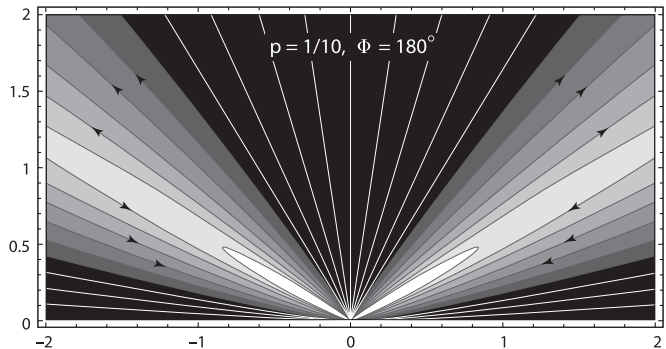}
\caption{After half a turn a field stretches to very large distances
  in the absence of a confining coronal pressure.  After ${2\pi \over
  {\sqrt 3}}\ {\rm turns}\ = 207\deg$ the field stretches to infinity
  and further turning does no work.  For more recent work on field
  opening see \citet{uzab} and \citet{uz02}.}
\end{center}
\end{figure}

\noindent
Years later I wrote up
this problem \citep{dlb94} for a conference celebrating Mestel's work.
It proved to be a fine example of a phenomenon long advocated by
\citet{aly8495} in solar MHD, see also \citet{stu91}.  This rekindled
my interest, but it still took two years for me to realise 
an ambient coronal pressure external to the field would
prevent it expanding to infinity.  To do so would take too much
energy.  When the field cannot expand to infinity, the continual
turning of the accretion disk does wind many turns into the corona,
provided the conductivities are great enough for flux freezing at the
feet.  Then the original argument comes into operation and tall towers
will be generated whose heights grow with each successive turn.  Basic
theorems on magneto-statics and much simplified models of such
magnetic towers in a constant external pressure were given in paper II
\citep{dlb96} which used a rigidly rotating inner disk and a fixed
outer disk to which the field returned.  Simplified models with
realistically rotating accretion disks were derived in a conference
paper \citet{dlb01}.  The present paper is a development of those
calculations to allow for a pressure that decreases with height.  We
also find a better approximation for the distribution of twist with
height along each field line.  This improves on the assumption of a
uniform distribution made earlier.

We derive the {\bf magneto-statics} of force--free magnetic fields whose
feet have been twisted by an accretion disk.  We study the {\bf
  equilibria} as a function of the twist angles.  The magnetic field
lines labelled $P$ being twisted by an angle $\Phi(P)=\Omega(P)t$ at
their feet which are anchored in the disk.  Here $\Omega(P)t$ is the
differential rotation of the feet of the flux tube and $t$ is a
parameter that gives the amount of that twist.  If we increase $t$ we
change our equilibrium along a Poincar\'{e} sequence.  Thus, in the
language of his catastrophe theory, $t$ is our control parameter.  Of
course, if all inertia were unimportant and if a real accretion disk
were dragging the feet of the flux tubes at relative angular velocity
$\Omega(P)$, then this sequence of equilibria would give us a film of
how the field would evolve in the presence of the external pressure
field $p(z)$.  This would not give a true picture just because the
inertial terms are not always negligible, nevertheless such sequences
are very instructive in that they show us how the field would like to
change if inertia did not slow its accelerations.  We find it possible
to get a reasonable understanding of this problem for any specified
$\Omega(P)$ and any chosen external pressure field $p(z)$.  We
emphasise that we have the simplest magnetically dominated model.
Above the disk, magnetism so dominates that the field is force--free
everywhere within the towers, but at their surfaces the magnetic
forces balance the ambient hot gas pressure $p(z)$.  Where there is
gas there is no field, where there is magnetic field there is no gas
pressure but we nevertheless assume perfect electrical conductivity.
When we view our sequences of equilibria parameterised by $t$ they are
so reminiscent of what is seen in radio galaxies, quasars,
star--forming disks with Herbig--Haro objects, etc., that it is hard
to resist the temptation of talking in the terminology of dynamics
rather than statics.  Please remember that the velocities we talk of
are only  the velocities that would occur in the absence of inertia
when accelerations can become arbitrarily large without penalty.  We
consider some effects of inertia in paper IV \citep{dlb02}.  

We believe that the very simple calculation given in section \ref{sec2.3}
contains the answer to the question `Why do flat accretion disks
produce needle like magnetic jets?', but further refinements, given in
paper IV on detailed field structure, give greater understanding of
why jets act as giant linear accelerators and why their electric
currents are so concentrated to the tower's axis.  

In the above, I have given a personal account of the evolution of my
thoughts and calculations but there are many other ideas in this
field, which started with the Jet in M87 observed by \citet{cur18}.
Although \citet{sch12} wrote a book which detailed the theory of
synchrotron radiations in 1912, it was not applied to M87 until
\citet{baa56}.  Even though the first quasar 3C273 had a prominent jet
these remained an enigma emphasised by \citet{whe71} in 1970, who
rightly 
drew attention to a paper by \citet{leb70}; \citet{ree71} provided
one of the first explanations which evolved into \citet{bla74} and
\citet{sch74}.  \citet{bar72} and \citet{lov76} were among the first
to treat magnetically dominated models, while for black holes
\citet{bla77} gave such ideas a new and most interesting twist by
producing a magnetic mechanism for extracting the spin energy from a
black hole.  \citet{bla82} have a nice mechanism for driving winds
centrifugally.  Extra-galactic radio jets were reviewed with a
wonderful series of radio pictures by \citet{beg84}.   \citet{sak8587}
showed a slow asymptotic collimation of such winds and the wind
equations have been much studied \citep{hey89, app93, pud87, lov91}.
\citet{oka99} criticised the collimation claims of most workers.
Computational simulations of jets have been made by \citet{bel9596}
and \citet{ouy97, ouy99}.  Many of these studies start with a uniform
magnetic field at infinity as did \citet{lov76} and \citet{shi8586}
who produced very convincing jets this way.  However, a uniform field
at infinity puts in a collimation at the start, whereas some wish to
see both the field and the collimation to emerge as a consequence of
the persistent rotation rather than having them inserted as a boundary
condition.  
Jets are by no means confined to relativistic objects but
are common in the accretion discs that occur around young stars.  In
fact, not long after accretion disks were applied to quasars and
mini--quasars \citep{dlb6971}, and the galactic centre \citep{eke71,
  dlb72}, they were applied to star formation \citep{dlb74}.  The jets
that emerge from star--forming disks do not travel at relativistic
speeds but at speeds of one or two hundred km/s.  There is a
correlation between the maximum circular rotational velocity within
the object and the speed of the jet and we give an explanation of this
in sections \ref{sec2.3} and \ref{sec3.3}.  First numerical
calculations of the force--free structures we are advocating, were
made by \citet{li01} but to date they have only calculated the first
two turns which give indications of the initial growth of magnetic
towers.  \citet{kra99} studied the launching of jets loaded with
plasma, as did \citet{rom98}, \citet{con95a} has - like us - looked
for force--free self--similar solutions for jets and has considered
the possibility of purely toroidal fields \citep{con95b}, Winds and
Jets are considered by \citet{lov91} and Poynting jets from loaded
winds  have been computed by \citet{lov02}. 

\section{Magneto-static Theorems and Deductions}\label{sec2}
\subsection{Theorems}\label{sec2.1}
We consider force-free magnetic configurations with the feet of the
flux tubes anchored in the accretion disk at $z=0$ and the magnetic
configuration confined by an external coronal gas pressure $p(z)$
which may depend on height $z$.  Where there is magnetic field, there
is no gas pressure.  Earlier versions of these theorems were proved
only for the constant pressure case in paper II.

The work done against the pressure in making a magnetic cavity whose
cross-sectional area at height $z$ is $A(z)$ is 
\begin{equation}
W_{p}= \int p(z) A(z) dz\ . \label{eq1}
\end{equation}
This may be thought of as a potential energy.  The larger the cavity
occupied by magnetic field the greater is the $pV$ energy stored, an
asset that could be drawn upon in recession by contraction.  The total
potential energy of the configuration stored in both magnetic energy
and $W_p$ is 
\begin{eqnarray}
W & = & 8\pi^{-1}\int\left (B_x^{2}+B_y^{2}+B_z^{2}\right )
dV+W_p\label{eq2}\\
& = & W_{x}+W_{y}+W_{z}+W_p.\nonumber
\end{eqnarray}
Contributions to these energies from slices at different heights $z$
are given by
\begin{displaymath}
w_{x}=(8\pi)^{-1}\int \int B_{x}^{2} dx dy\ ,
\end{displaymath}
over the area $A$, etc and 
\begin{displaymath}
w_{p}=p(z)\, A(z)\ .
\end{displaymath}
Evidently, $W_{x}=\int w_{x}dz,\ {etc, and}\ W_p =
\int w_{p} \, dz.$ 
The minima of the potential energy give stable equilibrium
configurations.  At any equilibrium (stable or not) the energy is
stationary, so the work done by any {\bf small} displacement
consistent with the constraints is zero.  We consider a small vertical
displacement caused by expanding the slice of the configuration
between $z$ and $z+\delta z$ (see Figure 2) so that it now occupies the
region between $z$ and $z+\mu \delta z$.  The region below $z$ is
unchanged; the region above $z + dz$ is lifted by $(\mu -1)\delta z$.
The work done by the pressure at the edge of the slice is
\begin{displaymath}
{\scriptscriptstyle{\frac{1}{2}}} 
(\mu-1)\delta z\delta z(-dA/{dz})p.
\end{displaymath}
which is second order in $\delta z$ and does not concern us; but
significant work is done on the pressure elsewhere because the area
$A(z^{\prime})$ that was initially at $z^{\prime}$ has been moved to 
$z^{\prime} +
(\mu-1)\,\delta z$, so the area at $z^{\prime}$ is now
\begin{displaymath}
A-(\mu-1)\delta z dA/dz^{\prime}\ .
\end{displaymath}  
Thus the change in $W_p$ due to the displacement is 
\begin{eqnarray*}
\Delta W_p \, & = &(\mu-1) \delta z
\int^{\infty}_{z}\,p(z^{\prime})(-dA/dz^{\prime}) 
dz^{\prime}\ .\\
& = &(\mu -1) \delta z  \left [p(z) \,  A(z)+ \int ^{\infty} _{z} \, A(z
^{\prime})\,(dp/dz^{\prime})\, dz^{\prime}\right ]\ .
\end{eqnarray*}
To get the second equality we integrated by parts and used the fact
that $A$ vanishes at sufficiently great heights.  This expression does
indeed yield $pdV$ when the pressure is constant.

\begin{figure}
\begin{center}
\includegraphics[scale=0.50]{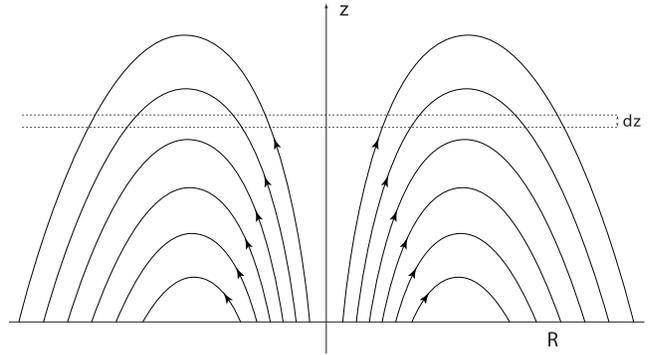}
\caption{A slice of a force--free field structure.  The twists about
  the axis are not shown.  The theorem is true without assumptions of
  axial symmetry.  To prove theorem I we consider a virtual
  displacement in which the slice $dz$ is expanded by a factor of
  $\mu$.  At equilibrium an infinitesimal expansion will do no work.}
\end{center}
\end{figure}

Magnetic fluxes must be conserved during the displacement so 
\begin{displaymath}
B_{x} \rightarrow \mu ^{-1} B_x , \ B_y 
\rightarrow \mu^{-1}B_y , \, B_z
\rightarrow B_z\ .
\end{displaymath}
within the expanded slice whose volume increases by a factor
$\mu.$  Evidently
\begin{eqnarray*}
8\pi \Delta W_x & = & \mu ^{-2}\int \int B^2_x dxdy \mu \delta z - 
\int \int B_x^{2} dxdy \delta z\\
 & = & (\mu ^{-1} -1) \int \int B_x^{2} dxdy \delta z\\
8 \pi \Delta W_z & = & \, (\mu-1) \int \int B_z^{2} dxdy  \delta z\ .
\end{eqnarray*}
Collecting all the $\Delta W$ terms dividing by
$(\mu -1) \, \delta z\ 
\textrm{and letting}\ \mu \rightarrow 1$
we have theorem I.
\begin{equation}
-(w _{x}+w_{y}) + w_{z} + p(z)A(z) + \int^{\infty}_{z} 
A(z^{\prime}) dp/dz^{\prime} dz^{\prime} =0\ . \label{eq3}
\end{equation}
If we integrate over all $z$  (by parts for the last term), we obtain
\begin{equation}
-(W_x+W_y) + W_z + W_p + \int^{\infty}_{0} z A(z)
 dp/dz dz = 0 \ .\label{eq4}
\end{equation}
The integrals in both (\ref{eq3}) and (\ref{eq4}) are negative when $p$
decreases with 
height but vanish when $p$ is constant.  If $d{\rm ln}p/d{\rm ln}z=-s$,
then the final integral in (\ref{eq4}) is  $-sW_p$.
Our next theorem is obtained by taking our virtual displacement to be
a uniform lateral expansion with no vertical shift.  Here we can not
apply the virtual work principle unchanged since one of the
constraints is that the vertical magnetic flux is frozen at the anchor
points all over the disk $z=0.$  Our lateral expansion would violate
this.  However, there is an extension of the principle in which the
constraints are replaced by the forces of constraint in the
equilibrium configuration and the virtual work then contains a
contribution from the work done against the forces of constraint.  In
our case these forces are those due to the disk that balance the
equilibrium Maxwell stresses.
\begin{displaymath}
-(4 \pi)^{-1} \left [B_{R} B_{z} {\bf {\hat{R}}} + 
 B_{\phi} B_{z} 
 \mbox{\boldmath$\hat\Phi$} 
+  \, {\scriptscriptstyle \frac{1}{2}} \left(B_{z}^{2}-B^{2}_{R}-B^{2}_{\phi} \right) 
{\bf{\hat{z}}} \right]
\end{displaymath}
per unit area of disk.  The work done on the disk in a
radial displacement 
$\mbox{\boldmath$\xi$} = \mu {\bf R}$ is
\begin{displaymath}
\mu W_0 = (4\pi)^{-1} \mu \int\int B_R B_z \ \ R^2d\phi dR
\end{displaymath}
evaluated in $z=0$.

The change in $W_p$ is just $(\mu^2-1)\int pAdz$ since \linebreak$A\rightarrow
\mu^2A$, similarly $B_x \rightarrow \mu^{-1}B_x$, $B_y \rightarrow
\mu^{-1}B_y$ and $B_z \rightarrow \mu^{-2} B_z$.  Since $dV\rightarrow
\mu^2dV$ we find
\begin{displaymath}
\Delta W = \left (\mu^{-2}-1\right) W_z + \left (\mu^2 -1\right ) W_p
\end{displaymath}
with NO contribution from $W_x $ or $W_y$ since they are unchanged.  The
condition of equilibrium is no longer $\Delta W  = 0$ but rather that any
decrease in $W$ must be balanced by the work done against the forces
of constraint i.e., 
\begin{displaymath}
-\left [d\Delta W/d\mu \right ]_{\mu=1} = W_0\ .
\end{displaymath}
Thus we deduce Theorem II
\begin{equation}
W_z - W_p = {\scriptscriptstyle{{1\over 2}}} W_0\ . \label{eq5}
\end{equation}
Many scientists brought up on the pinch effect find it strange that
the toroidal component of the magnetic field has NO effect on the
change of magnetic energy in such a global lateral expansion.  In
fact, there is no general tendency for the whole magnetic structure to
shrink towards an axis.  In axial symmetry $W_x + W_y = W_R + W_\phi$
and none of them changes in a {\bf uniform} contraction transverse to
the axis.  We return to explain the relationship of this result to the
pinch effect in section \ref{sec2.5}.  

If, in place of the lateral expansion at all heights, we freeze the
configuration below some height $z$ and laterally expand everything
above that height, then it is the magnetic stresses across the plane
at height $z$ that become our new $W_0$ so we call their contribution
$W_0(z)$.  Likewise, the $W_z$ and $W_p$ involved are integrated over
the region above height $z$ only.  We shall call these quantities
$W_z(z)$ and $W_p(z)$.  We may then generalise (\ref{eq5}) to give the
exact result (\ref{eq5*})
\begin{equation}
W_z(z) - W_p(z) = {\scriptscriptstyle{1\over 2}} W_0(z) = {1\over
  8\pi}\int\int_z B_RB_zR^2d\phi dR\ . \label{eq5*}
\end{equation}

To localise 
theorem II 
consider a uniform lateral expansion that
varies slowly with $z$ so that ${\bf R} \rightarrow \left [1 +
  \mu (z)\right ]{\bf R}$.  We shall take $\mu$ to be zero in $z=0$
and to climb to a maximum at $z_1$ before declining again to zero.
Provided $\mu(z)$ varies slowly enough the shear $\mu(z)$ will not
produce significant radial field from the displacement of vertical
field.  If we neglect this effect, the scalings are as for theorem II
and we get a version of it localised near $z_1$ (which we then replace
by $z$).  Then our theorem III reads for $z/R \gg 1$ and $dB/dz \ll
\vert B\vert/R$
\begin{equation}
w_z \simeq  w_p \label{eq6} 
\end{equation}
$W_0$ does not contribute as there is no displacement on $z=0$.  
(\ref{eq6}) may also be derived by differentiating the result
(\ref{eq5*}) with respect to $z$.  That derivation demonstrates that
(\ref{eq6}) is valid provided $d/dz\left ({\scriptscriptstyle{1\over
      2}}W_0(z)\right )$, is small compared with $w_z$, which is so
  provided $B_R \ll B_z$ at height $z$.
Unlike
theorems I and II, theorem III is approximate and only true well away
from the disk.  If we integrated it down to the disk it would conflict
with theorem II except in the special case when $W_0$ is zero.  
\subsection{Winding Makes Tall Towers}\label{sec2.2} 
We shall use cylindrical polar coordinates but we shall not assume
axial symmetry.  Then theorem I may be written
\begin{displaymath}
W_R + W_\phi = W_z +(1-s)W_p
\end{displaymath}
adding $W_z + W_p$ and using (theorem II)
\begin{eqnarray*}
W=W_R+W_\phi+W_z+W_p & = & (4-s) W_p + W_0\\
      & = &  (4-s) W_z+{\scriptscriptstyle{{1\over2}}}(s-2)W_0\ .
\end{eqnarray*}

Remember that for the constant pressure case $s=0$.\\
Here we shall assume $\langle-d{\rm ln} p/d{\rm ln}z\rangle =s<4$.  
We shall show elsewhere that when 
  $\langle-d{\rm ln} p/d{\rm ln}z\rangle \geq 4$
 the
magnetic field balloons off to infinity as in the highly wound
pressureless case of paper I  We assume that even after several turns
the field structure resists being wound still further and that the
work done per turn is asymptotically a constant.  Furthermore, we
shall assume that the boundary term $W_0$ which only involves the
field on $z=0$ tends to a limiting value.  Then in each turn $W$ will
be raised by a finite $\Delta W $ and $\Delta W = (4-s)\Delta W_p =
(4-s)\Delta W_z$.  Hence $W_p$ and $W_z$ must increase without limit
as the winding continues.  Now $W_p = \int p(z)A(z)dz$ can increase by
increasing $A(z)$ at given $z$ or by increasing the height to which
the whole configuration reaches.  However, increase of $A(z)$ for
given $z$ does not increase $W_z$ since $B_z$ is of order $F/A(z)$
where $F$ is the poloidal flux.  Thus $\int B^2_z dxdy \propto
F^2/A(z)$.  Since $W_p$ and $W_z$ have to increase together we deduce
that the height of the whole structure increases for each turn of the
flux anchor points on the disk.  This argument is reinforced by the
very crude estimates of field structure that follow and by the more
accurate but more specific field models calculated later. 
When the configurations are tall, certain simplifications occur in
the theorems.  Defining $\langle B^2_z \rangle = A^{-1}\int\int B^2_z
dxdy$ theorem III becomes

\begin{equation}
\langle B^2_z \rangle = 8\pi p(z)\ , \label{eq8}
\end{equation}
where the average is taken at height $z$.  Secondly, there is only a
finite poloidal flux and as the system gets taller and taller the
radial flux through a cylinder gets more and more spread out.  Thus
$B_R$ becomes very small compared with $B_\phi$ and $B_z$.  If $W_R$
is neglected in theorem I, we have 
\begin{equation}
w_\phi = w_z + w_p + \int^\infty_z A(z^\prime) dp/dz^\prime dz^\prime\
. \label{eq9} 
\end{equation}
In the pressure constant case we see from (\ref{eq8}) and (\ref{eq9}) 
\begin{displaymath}
w_\phi = 2w_z = 2w_p 
\end{displaymath}
so
\begin{equation}
\langle B^2_\phi \rangle = 2 \langle B^2_z \rangle = 16 \pi p \
. \label{eq10}
\end{equation}
When $p$ varies we define 
\begin{displaymath}
\sigma(z) = w^{-1}_p \int^\infty_z A(z^\prime) \left (-dp/dz^\prime
\right ) dz^\prime  \ , 
\end{displaymath}
then  
we have 
\begin{equation}
\langle B^2_\phi \rangle = \left (2-\sigma\right ) \langle B^2_z
\rangle = \left (2-\sigma \right ) 8\pi p \ , \label{eq11} 
\end{equation}
When the vertical Maxwell stresses $(8\pi)^{-1} \left (B^2_z -
  B^2_\phi - B^2_R \right ) $ are integrated over a cross section we
  find a net tension of $w_z - w_\phi - w_R$ which becomes a force
  driving longitudinal expansion when that quantity is negative as in
  (\ref{eq3}), (\ref{eq9}) or (\ref{eq11}).
\subsection{Crude Estimates Give Essential  Understanding}\label{sec2.3}
 Let the total poloidal magnetic field emerging from an accretion disk
 on $z=0$ be $F$.  We shall assume that it returns to the disk at some
 larger radius so that it is anchored at both ends.  Suppose that a
 typical field line reaches a height $Z$ and that the tall magnetic
 structure has a radius $R$.  Now each turn of the poloidal flux
 generates an equal toroidal flux which must pass through the area
 $RZ$.  Hence after $N$ turns of the feet relative to one another the
 typical $B_\phi$ is of order $NF/RZ$ and the volume is $\pi R^2Z$ so 
\begin{displaymath}
8\pi W_\phi = N^2F^2\pi/Z\ .
\end{displaymath}
The vertical flux $F$ goes once up and once down so if it goes up in
the inner $R/\sqrt{2}$ and down outside that $\vert B_z\vert \simeq
2F/\pi R^2$ and have
\begin{displaymath}
8\pi W_z = 4F^2Z/(\pi R^2)
\end{displaymath}
Finally the flux passes through a cylinder of radius $R/\sqrt{2}$
radially so 
\begin{displaymath}
B_R = F/\left (\sqrt{2}\pi RZ \right ) \ ,
\end{displaymath}
and hence
\begin{displaymath}
8\pi W_R = F^2/\left (2\pi Z \right ) \ .
\end{displaymath}
For a constant pressure $p$ we have 
\begin{displaymath}
W_p = p\pi R^2 Z \ .
\end{displaymath}
Hence
\begin{equation}
W={F^2 \over \pi} 
\left [ \left (
N^2 \pi^2 + {\scriptscriptstyle{1\over2}}\right )
/Z + 
\left ( 
4R^{-2} + 8\pi^3 pF^{-2}R^2 \right ) Z \right ]\ . \label{eq12}
\end{equation}
Minimising $W$ over all values of $R^2$ we find
\begin{equation}
\pi R^2 = F \left (2\pi p \right )^{\scriptscriptstyle{1\over 2}}
 \ , \label{eq13}
\end{equation}
so the area is determined by $F$ and $p$ whatever the minimising $Z$
may be and the two terms that make up the coefficient of $Z$ are equal
at equilibrium.  Minimising $W$ over $Z$ we find
\begin{displaymath}
Z = N\pi \left (R/\sqrt{8}\right ) \ 
\left (1+{\scriptscriptstyle{{1\over 2}}} N^{-2}\pi^{-2}
\right )^{\scriptscriptstyle{{1\over 2}}}
\end{displaymath}
with $R$  already given by (\ref{eq13}) independent of $N$, this
clearly shows that the height $Z$ increases linearly with $N \gg 1$.
Indeed, if we write $N=\left (2\pi\right)^{-1}\Omega t$ where $\Omega$ is
the relative angular velocity of the flux feet, then
\begin{equation}
Z\rightarrow {1\over 4\sqrt{2}} \Omega Rt \ , \label{eq14}
\end{equation}
so the height of a steadily wound magnetic structure will grow with a
velocity $\propto \Omega R$.  Notice that $R$ is typically larger than the $R$
of the inner foot of the flux tube at whose radius $\Omega $ is determined.
Thus even with this excessively crude 
model we can see why the velocities of growing magnetic towers are
directly related to the velocities in the accretion disk.  

In this section we have assumed thus far that our growing towers do
not have hollow cores with no field in them; consider however the top
of a tower on axis.  There the magnetic field must  splay out radially
before descending down the exterior of the tower.  Normally such
division of one field line only occurs at a neutral point but here the
magnetic field must resist the external pressure $p$ so $B^2/8\pi$
cannot be zero.  Thus on the axis there has to be a most interesting
cusp point.  The cusp points downward and has a vanishing opening
angle.  The axial field line comes to the cusp at finite field
strength and there splays very gradually at first, but then more
rapidly with the field strength $B^2/8\pi$ balancing the `external'
pressure which has invaded the axis and its neighbourhood, down to the
level of the cusp.  For points above such a cusp the tower will be
hollow.  Of course, it may be that this cusp is a mere dimple in the
top of the tower, however the MHD calculations of \citet{li01}, who
were only able to calculate the towers for the first two turns,
suggest that the towers may be hollow over a significant fraction of
their height.  Let us recalculate the energy assuming that our tower
has 
a hollow core radius of 
$R_c$
and a maximum radius $R_m$. Following our previous calculations 
\begin{displaymath}
B_\phi = NF/\left [\left (R_m-R_c \right )Z\right ]\ ,\\
\end{displaymath}
\begin{displaymath}
8\pi W_\phi = N^2F^2\pi \left [ \left
    (R_m+R_c\right)/\left(R_m-R_c\right )\right ]/Z\ .
\end{displaymath}
\begin{displaymath}
\vert B_z\vert = 2F/\left [\pi \left (R^2_m - R^2_c \right
  )\right ]\ ,\\
\end{displaymath}
\begin{displaymath}
8\pi W_z = 4F^2Z/\left [\pi \left (R^2_m - R^2_c \right )\right ] \ .\\
\end{displaymath}
\begin{displaymath}
B_R = F/\left \{2\pi \left [{\scriptscriptstyle{1\over2}}\left (R^2_m
      + R^2_c \right )\right ]^{\scriptscriptstyle{{1\over 2}}}
Z\right \} \ ,\\
\end{displaymath}
\begin{displaymath}
8\pi W_R = \left [F^2/\left (2\pi Z \right )\right ] \left (R^2_m - R^2_c
\right )/\left (R^2_m+R^2_c\right )\ , \\
\end{displaymath}
\begin{displaymath}
8\pi W_p = 8\pi^2 p \left (R^2_m - R^2_c \right ) Z\ .
\end{displaymath}

Writing $x=R_c/R_m$ we deduce

\begin{equation}
\begin{array}{l}
W=F\pi^{-1}
\end{array}
\left \{
\begin{array}{l}
\left [
N^2\pi^2 
{\left (1+x\right )\over 
\left (1-x\right ) }
+ {\scriptscriptstyle{1\over 2}}
{\left (1-x^2\right )\over \left(1+x^2\right )}
\right ] Z^{-1} +\\
+ \left [
4\left(1-x^2 \right )^{-1}
R^{-2}_m + \right.\\ 
\hfill\left. + 8\pi^3 p F^{-2} R^2_m
\left (1-x^2 \right )
\right ]
 Z \ 
\end{array} \right \} \ . \label{eq15}
\end{equation}
Minimising over $R^2_m$ keeping $x$ and $Z$ fixed we find 
\begin{displaymath}
R^2_m = \left (2\pi^3 p\right )^{-{\scriptscriptstyle{{1\over 2}}}} 
F \left (1-x^2 \right )^{-1}
\end{displaymath}
so at these minima we have for all  $x$ and $Z$ 
\begin{displaymath}
W_0 = F^2\pi^{-1}
\left \{
\begin{array}{l}
\left [N^2\pi^2 
{\left (1+x \right )\over 
\left (1-x\right )} + {\scriptscriptstyle{1\over 2}}
{\left (1-x^2\right )
\over \left (1+x^2\right )} \right ] Z^{-1} + \\
\hfill +8\left (2\pi^3p\right )^{{\scriptscriptstyle{{1\over 2}}}} F^{-1}Z
\end{array}
\right \} \ .
\end{displaymath}
Minimising over all $x$ in the range $0\leq x <1$ we find that the
quantity in square brackets increases throughout the range so the
least value occurs at $x=R_c/R_m =0$.
Thus, at the level of this crude approximation, the towers are not
hollow but filled with magnetic field.
\subsection{Stability}\label{sec2.4} 
Potential energy minima are always stable but if the minimisation is
carried out under the restriction of axial symmetry there may exist
asymmetrical distortions that lead to lower energy configurations.  A
useful insight was taught me by Moreno Insertis.  Magnetic towers
in tension are stable to sideways bowing.  Like Euler struts, tall
towers in compression are unstable.  Such ideas require modification
in the presence of an external pressure.  The tower of fluid giving
the ambient pressure is not unstable although it is under
compression.  Thus a magnetic tower that merely gives ambient pressure
will not be unstable this way, rather it is pressures above ambient
that cause such instabilities.  The modified stability criterion is
$w_\phi - w_z \leq w_p$.  This criterion is only marginally satisfied
by our constant pressure configurations for which the equality holds,
so those will be pretty floppy, but when $p$ decreases with height the
buoyancy term provides stability.   
\subsection{The Pinch Effect a Pressure Amplifier}\label{sec2.5} 
By showing that the energy in the $B_\phi$ and $B_R$ components of the
field did not change in a uniform lateral expansion we demonstrated
that there is NO tendency for a highly wound magnetic structure to
contract overall laterally.  Where then does the Pinch Effect come
from?  To elucidate this consider a specified toroidal magnetic flux
$F_\phi$ contained between inner and outer cylinders of radii $R_i$
and $R_0$ both of height $Z$.  Minimising the total energy over all
possible $B_\phi(R)$ that give the flux, the minimum occurs with
$B_\phi \propto R^{-1}$ and the energy is then
\begin{displaymath}
W_\phi = {\scriptscriptstyle{1\over 4}}F^2_\phi Z^{-1}/\left [{\rm
    ln}\left (R_0/R_i \right )\right ]\ .
\end{displaymath}
 In uniform lateral contractions or expansions $R_0/R_i$ remains
 constant, but if we fix $R_0$ then $W_\phi$ can be decreased by making
 $R_i$ smaller. 

It is this that gives the pinch effect.  Since $B^2_\phi\propto
R^{-2}$ the toroidal magnetic field acts as a pressure amplifier
delivering on the cylinder at $R_i$ a pressure $R^2_0/R^2_i$ times the
pressure it exerts on the outer cylinder.  The whole pinch effect
fails if there is nothing to push on at the outside and indeed the
field would expand outward to larger radii reducing the pressure on
$R_i$ to zero.  The Pinch Effect is a {\bf pressure amplifier};
amplifying nothing gives nothing.  In the presence of a $B_z$ the
amplifier has less gain.  By our theorems $(8\pi)^{-1} \int B^2_z 2\pi
RdR = \pi \left (R^2_0 p_0 -R^2_i p_i \right )$ and if $(8\pi)^{-1}\int
B^2_z 2\pi R dR = \eta \pi R^2_0 p_0$ then $p_i = (1-\eta)p_0
R^2_0/R^2_i$ so the amplifier only amplifies the excess pressure
unbalanced by $B^2_z$.  
\section{Confined Force--Free Magnetic Fields}\label{sec3} 
\subsection{Basic Equations}\label{sec3.1} 
Inside the tower the magnetic fields dominate over any other forces so
the magnetic field takes up a configuration that delivers no body force
inside the jet, i.e., a force--free configuration with ${\bf j}\times
{\bf B} = 0$.  Thus ${\bf j}$ is parallel to ${\bf B}$ and we may
write ${\bf j} = \widetilde{\alpha} {\bf B}$ where $\widetilde{\alpha}$ is a
  scalar function of position.  Now both ${\bf j}$ and ${\bf B}$ have
  no divergence so we deduce that ${\bf B} \cdot
  \mbox{\boldmath$\nabla$} \widetilde{\alpha} = 0$ which implies that
  $\widetilde{\alpha}$ is constant along each line of force.  We consider
  axially symmetrical systems in cylindrical polar coordinates
  $(R,\phi,z)$ in terms of the poloidal flux function $P(R,z)$
  which gives the flux through a circle radius $R$ at height $z$.  By
  flux conservation

\begin{equation}
B_z=(2\pi R)^{-1}\partial P/\partial R \ , \label{eq16}
\end{equation}
and
\begin{equation}
B_R = - (2\pi R)^{-1}\partial P/dz\ . \label{eq17}
\end{equation}
We write
\begin{equation}
B_\phi = (2\pi R)^{-1}\beta \ , \label{eq18}
\end{equation}
and then deduce
\begin{equation}
{\bf B}= \nabla P \times \nabla (\phi/2\pi)+\beta \nabla (\phi/2\pi)\
, \label{eq19}
\end{equation}
$\beta$ is an axially symmetrical scalar function of position.
Evaluating the curl of (\ref{eq19}) we find
\begin{eqnarray}
4\pi {\bf j} =  \nabla \times {\bf B} = \nonumber \\
- \left[R \partial/\partial R \left (R^{-1}\partial P/\partial R
  \right ) + \partial^2 P/\partial z^2 \right ]\nabla 
\left (\phi/2\pi\right ) + \nonumber \\
+ \nabla \beta \times \nabla \phi/2\pi \ . \label{eq20}
\end{eqnarray}
The force-free condition ${\bf j}= \widetilde{\alpha} {\bf B}$ now
gives, cf (\ref{eq19}), both 
\begin{equation}
\nabla \beta = 4\pi \widetilde{\alpha}\nabla P\ , \label{eq21}
\end{equation}
and
\begin{equation}
R \partial/\partial R (R^{-1} \partial P/\partial R ) + \partial^2
P/\partial z^2 = - 4\pi \widetilde{\alpha}\beta \ . \label{eq22}
\end{equation}
From (\ref{eq21}) it follows that the normals to the surfaces of
constant $\beta$ are the normals to the surfaces of constant $P$ so
$\beta$ is a function of $P$ and 
\begin{equation}
\beta^\prime(P) = 4\pi\widetilde{\alpha}\ . \label{eq23}
\end{equation}
Inserting this expression for $\widetilde{\alpha}$ into (\ref{eq22}) we get
the basic equation for force--free equilibrium (we write $Q(P)$ for
$\beta\beta^\prime$) 
\begin{equation}
R \partial/\partial R(R^{-1}\partial P/\partial
R)+\partial^2P/\partial z^2 = -\beta\beta^{\prime} = -Q(P) \
. \label{eq24}
\end{equation}
From (\ref{eq19}) we see that ${\bf B}\cdot\mbox{\boldmath$\nabla$} P =
  0$ so that $P$ is constant along a line of force.  Indeed the
  equations for the lines of force follow from the condition $d{\bf s}
  \vert\vert {\bf B}$ which gives
\begin{displaymath}
{dR \over B_R} = {Rd\phi \over B_\phi} = {dz \over B_z}\ ,
\end{displaymath}
or using the above expressions for ${\bf B}$
\begin{equation}
-{dR \over \partial P/\partial z} = {Rd\phi \over \beta(P)} = {dz
 \over \partial P/\partial R} . \label{eq25}
\end{equation}
We shall use this later to work out the twists of field lines.
\subsection{Field Structure Above a Differentially Rotating
  Disk}\label{sec3.2} 
Consider the tube of flux that rises within an inner circle of radius
$R_i(P)$ on an accretion disk and returns to it at some outer radius
$R_0(P)$.  There will be a differential twisting due to the fact that
$\Omega_i >\Omega_0$ and both are radius dependent.  We define
$\Omega(P) = \Omega_i - \Omega_0$ so $\Omega(P)$ is the rate of
differential twisting of the field lines labelled $P$.  At any time
$t$ when the twist angle has accumulated to be $\Phi(P)=\Omega(P)t$,
the field lines labelled $P$ will rise to some maximum height $Z(P)$
before heading down to re-intersect the disk at $R_0(P)$.  We are
concerned to know how this total twist $\Phi$ is distributed over the
height $z$.  In \citet{dlb01} I found that great progress could be
made by simply adopting the idea that each field line had a certain
twist per unit height $\Phi(P)/Z(P)$, however I did not use the full
power of the variational principle to derive the equations.  Equations
(\ref{eq25}) for the lines of force show that 
\begin{equation}
d\phi/dz = \beta(P)/(R\partial P/\partial R) \ , \label{eq26}
\end{equation}
whereas $P$ and $\beta(P)$ are constant along each line, $R\partial
P/\partial R$ is not.  Now as the magnetic tower grows, its
cross--section at any height will have some dimensionless profile $f$
with 
\begin{equation}
P=P_m(z)f(\lambda)\ . \label{eq27}
\end{equation}
Here $P_m(z)$ is the maximum value $P$ achieves on a cross section of
height $z$.  Now $Z\left(P_m(z)\right)=z$, so $ P_m(z)$ is the
inverse function of $Z(P_m)$.  $\lambda$ is the fractional area of the
cross section at height $z$, $\lambda = \pi R^2/A(z)=R^2/R^2_m$ where
$R_m(z)$ is the radius of the magnetic cavity at height $z$ and $A(z)$
is its area.  
From these definitions the maximum value of $f$ is one and
since $P$ is zero on axis\footnote{If the tower were hollow $P$ would
  be zero for a whole area near the axis.} and at the tower's surface
$f$ will be zero at $\lambda = 0$ and 1.  If $f$ is independent of
height the tower is said to be self--similar, however, in general the
form of the function $f(\lambda)$ may depend on height although such
changes may be slow except at the tower tops and near their feet.
From (\ref{eq27}) 
\begin{displaymath}
R \partial P/\partial R = P_m(z) 2\lambda f^\prime(\lambda) =
2P/(d{\rm ln}\lambda/d{\rm ln}f)\ .
 \end{displaymath}
Thus from (\ref{eq26}) 
\begin{displaymath}
d\phi/dz={\scriptscriptstyle{1\over 2}}
P^{-1}\beta(P)  (d{\rm ln}\lambda/d{\rm ln}f)\ .    
\end{displaymath}
Within any small height interval $dz$ the twist of the line for force
labelled $P$ will have two contributions, one ${\scriptscriptstyle{1\over 2}} 
P^{-1}\beta(P)(d{\rm ln}\lambda_i/d{\rm ln}f)dz$ as the flux rises
through $z$ at $\lambda_i$
and a second $-{\scriptscriptstyle{1\over 2}} P^{-1}\beta (d{\rm
   ln}\lambda_0/d{\rm ln}f)dz$ as line descends through $z$ at
   $\lambda_0$.  
   Thus the total contribution to the twist
of the line from this height interval will be 
\begin{equation}
d\Phi = 
{\scriptscriptstyle{1\over 2}} 
P^{-1} \beta (P)
\left[{d{\rm ln}
       \left(\lambda_i/\lambda_0
       \right)\over d{\rm ln}f} 
\right]dz\ = 
  \left(
       \left.{d\phi\over dz}
       \right\vert_i - 
            \left.{d\phi \over dz}
            \right\vert_0
  \right) dz 
 , \label{eq28}
\end{equation}
as both $\lambda_i$ and $\lambda_0$ correspond to the same $P$ and are
at the same height, $z$, the value of $P/P_m(z) =f$ is the same for
each.  

Thus we may regard  $\lambda_i$ and $\lambda_0$ as the roots for
$\lambda$ of the equation $f(\lambda) = f$.  A simple example will
illustrate this.

Suppose $f$ is given by
\begin{equation} 
1/f = \lambda^2_1/\lambda + (1-\lambda_1)^2/(1-\lambda)\ .
\label{eq29}
\end{equation}
%
This $f$ is clearly zero at $\lambda = 0$ and $1$, we have
chosen the coefficients to ensure $f=1$ at the maximum at $\lambda
=\lambda_1$.  Now 
\begin{displaymath}
f^\prime = 
{\left (
    \lambda_1 -\lambda)(\lambda_1 +\lambda-2\lambda\lambda_1
 \right ) 
\over 
\left [\lambda^2_1 + \lambda(1-2\lambda_1)\right ]^2} \ ,
\end{displaymath}
and at $\lambda = \lambda_1$
\begin{displaymath}
f^{\prime\prime}=-2/\left [\lambda_1(1-\lambda_1)\right ]\ .
\end{displaymath}
Hence near the maximum of $f$ we find it is well approximated by
\begin{displaymath}
f=1-(\lambda-\lambda_1)^2 
\left [\lambda_1(1-\lambda_1)\right ]^{-1}\ ,
\end{displaymath}
which is actually exact for all $\lambda$ when $\lambda_1 = 
{\scriptscriptstyle{{1\over 2}}}$.
The roots for $\lambda$ are $\lambda_{0\atop i} = \lambda_1 \pm \left
  [\lambda_1 (1-\lambda_1)\right ]^{{\scriptscriptstyle{{1\over 2}}}}
(1-f)^{{\scriptscriptstyle{{1\over 2}}}}$ and 
\begin{equation} 
d{\rm ln}\left({\lambda_i\over\lambda_0}\right)
\left [{\rm ln}f = 
(1-f)^{{\scriptscriptstyle{{1\over 2}}}}
\left \{ 
{f \left [ \lambda_1(1-\lambda_1)\right ]^{{\scriptscriptstyle{1\over 2}}}\over
  \left [\lambda_1 -(1-\lambda_1)(1-f)\right ] }
\right \}\ \right]^{-1}  \label{eq30}
\end{equation}  
the curly bracket reduces to $1$ when $\lambda_1 =
{\scriptscriptstyle{1\over 2}} $.  
Near the maximum $1-f \ll 1$ so the expression becomes 
$(1-f)^{{\scriptscriptstyle{{1\over 2}}}}
(\lambda^{-1}_1-1)^{{\scriptscriptstyle{{1\over 2}}}} $.  
Away from the maximum $\lambda_i \simeq
\lambda^2_1 f \left [1+(1-\lambda_1)^2f\right ]$ and $\lambda_0 \simeq
1-(1-\lambda_1)^2f(1+\lambda_1f)$ so we find 
\begin{eqnarray*}
  d{\rm ln}\left({\lambda_i\over\lambda_0}\right)/d{\rm ln}f \simeq
\left [1-(1-\lambda_1)^2f\right ]^{-1} +\\
+ \left (1-\lambda_1\right )^2 f \left
  [1+(1-\lambda_1)^2f\right]^{-1}\ .
\end{eqnarray*}
For $\lambda_1 = {\scriptscriptstyle{{1\over 2}}} $ and $f={\scriptscriptstyle{{3\over 4}}}$ this gives ${19 \over 13}$ i.e.,
almost $1{1\over 2}$ but for $f = {1\over 2}$ it is ${70 \over 63}
\simeq 1 {1\over 4}$ and for $f = {1\over 4}$ it is $1{1\over 8}$.
Except for the integrable singularity of
$(1-f)^{-{\scriptscriptstyle{{1\over 2}}}}$   at the top of
each field line we deduce that 
in this simple example $d{\rm ln}(\lambda_0/\lambda_i)/d{\rm ln}f$
does not vary greatly.  The integrable singularity at the top of each
field line, $P = P_m$ is not a peculiarity of the example chosen.
Returning to the general case and looking near the maximum $f=1$ at
$\lambda = \lambda_1$ the roots $\lambda_0$ and $\lambda_i$ must
behave as $\lambda_1 \pm c \sqrt{1-f}$ so quite generally $d{\rm
  ln}(\lambda_i/\lambda_0)/d{\rm ln}f$ will contain the
$(1-f)^{-\scriptscriptstyle{{1\over 2}}}$ 
factor near $f=1$  We shall show in paper IV that in one limit
$f=-e\lambda {\rm ln}\lambda$ and in the opposite limit $f = 2\lambda$
or $2(1-\lambda)$ depending on whether $\lambda$ is less than or
greater than ${1\over 2}$.  In both these cases $d{\rm
  ln}(\lambda_2/\lambda_0)/d{\rm ln}f$ is near $1$ far from the top of
the field line and in the latter case it is $(1-f/2)^{-1}$ which
varies from 1 to 2 over the whole range $0\geq f \geq 1$. 

Notwithstanding such variations we shall start with the rough
approximation that twist is distributed uniformly with height so that
its distribution function 
\begin{equation} 
g(P,z)dz=dz/Z(P)\ . 
\label{eq31}
\end{equation} 
%
This integrates to 1 over the range $z=0$ to $Z(P)$ and gives so
simple a result that it may be used as a starting point of a refined
treatment that gives greater weight near the tops of field lines see
section 4.  With the distribution (\ref{eq31}) we calculate the
$B_\phi$ flux between $z$ and $z+dz$.  Each turn around the axis of an
element of poloidal flux $dP$ generates equal toroidal flux so the
twist $\Phi g(P,z)dz$, gives a toroidal flux
$(2\pi)^{-1}\Phi(P)g(P,z)dPdz$ in the element of height $dz$.  Adding
the contributions over all the $P$ that reach height $z$, i.e., those
with $P\geq P_m(z)$ we find that the total toroidal flux per unit $dz$
is 
\begin{equation}
\bar{B}_\phi R_m= \int B_\phi dR = \int^{P_m(z)}_0
(2\pi)^{-1}\Phi(P)g(P,z)dP\ . \label{eq32}
\end{equation}
We shall find it convenient to work in terms of average fields at a
given height, so the left--hand--side we re--write as $\bar{B_\phi}(z)R_m$.
Our former  $\left<B^2_\phi\right>$ will be related
to $\left (\bar{B_\phi}\right )^2$ by some structural constant related
to the profile of $B_\phi$ with $\lambda$.  We define the ratio $J^2$
by
\begin{equation}
J^2 = {<B^2_\phi >\over \left (\bar{B}_\phi\right )^2} = {4
\int\beta^2(P)\lambda^{-1} d\lambda \over \left
  [\int\beta(P)\lambda^{-1}d\lambda \right ]^2} \ . \label{eq33}
\end{equation}
Were the tower hollow for $R<R_c$ then multiply the right--hand--side
of (\ref{eq33}) by $(R_m -R_c)/(R_m + \vert \bar{B}_z\vert )$.
Similarly for the $z$ components we define $\vert B_z\vert$ cf
(\ref{eq16}), by
\begin{equation}
\vert{\bar{B_z}}\vert = A^{-1}\int^{R_m}2\pi R \vert B_z\vert dR =
2P_m/A \ , \label{eq34}
\end{equation}
and
\begin{equation}
I^2 = {\scriptscriptstyle{1\over 4}}\int \left({df \over
    d\lambda}\right)^2 d\lambda = {<B^2_z> \over \bar{\vert B_z
    \vert}^2} \geq 1 \ . \label{eq35}
\end{equation}
If the profile $f$ depends on height then $I^2$ and $J^2$ will in
general depend on height too, but for the tall towers generated by
continual winding we may expect that the profile form, $f$,  to settle
down to a typical one except near the tower top and bottom.  Thus $I$
and $J$ will not vary strongly and indeed for self--similar towers
they are dimensionless constants.  In paper IV we find for the
pressure constance case $I=1.359,\ J=.799$ and for the linear case
$I=1.179,\ J=1.098$.

Combining (\ref{eq8}) with (\ref{eq34}) and (\ref{eq35}) and setting
$R_c=0$
\begin{equation}
{P_m \over A} = {P_m\over \left (\pi R^2_m \right )} = 
{\scriptscriptstyle{1\over 2}} {<B^2_z>^{{\scriptscriptstyle{{1\over
          2}}}} \over I} =
{\left (2\pi p\right)^{{\scriptscriptstyle{{1\over 2}}}} \over I} \label{eq36}
\end{equation}  
in which all the symbols may be functions of $z$.

We use (\ref{eq36}) to write $R_m$ in terms of $P_m$ and $p$.  
Using (\ref{eq32}) to evaluate $\bar{B}_\phi$ via (\ref{eq30}),
(\ref{eq33}) gives us the average $B^2_\phi$ at given $z$
\begin{equation}
\langle B^2_\phi \rangle = J^2 \left [\int^{Pm}_0 (2\pi)^{-1}
  \Phi(P)/Z(P)dP\right ]^2R^{-2}_m\ . \label{eq37}
\end{equation}
$\langle B^2_z \rangle$ is given in terms of $P_m$ and $R_m$ by
(\ref{eq36}); while $B_R$ is so small that we could neglect $B ^2_R$
altogether, it is not uninteresting to see how its inclusion could
correct the result.  We shall make a very rough estimate.  The radial
flux up to height $z=Z$ through a cylinder of radius $R_1 =
R_m/\sqrt{2}$ will be $F-P_m(z)$.  This gives a mean radial field of
$\bar{B}_R = \left [F-P_m(z)\right/]\left (\sqrt{2}\pi R_m Z\right )$.
We therefore write
\begin{displaymath}
\langle B^2_R \rangle = \left [F-P_m(z)\right]^2/
\left (2 \pi^2R^2_m Z^2\right )\ .
\end{displaymath}
Substituting the above estimates into $8\pi W$ we have  $8\pi \bar{W}$ 
\begin{equation}
8\pi \bar{W} = \int{ 
\left \{
\begin{array}{l} 
I^{2}4\pi^{-1}P^2_m R^{-2}_m + 8\pi^2 p(z)R^2_m+\\
+ \pi J^2 \left [\int^{Pm}_0 (2\pi)^{-1}\Phi/Z dP \right ]^2 +\\
+ (2\pi)^{-1}(F-P_m)^2 Z^{-2}
\end{array}
\right \}}
dz \ . \label{eq38}
\end{equation}
Here $P_m(z) $ and $R_m(z)$ are functions of $z$ to be varied,
$Z(P_M)$ is the inverse function of $P_m(z)$ while $p(z)$ and
$\Phi(P)$ are given fixed functions.  $I,\ J,$ and $F$ we treat as fixed
constants. Only the first two terms involve $R^2_m$, varying it and
demanding that $\bar{W}$ be a minimum for all such variations gives
the two terms equal, but that equality merely reproduces what is
already contained in equation (\ref{eq36}) in the form $\pi R^2_m =
(2\pi p)^{-{\scriptscriptstyle{{1\over 2}}}}P_m {I}$.  
Inserting this $R^2_m$ into the first term and doubling the result as
the second is equal to it, those terms of $8\pi \bar{W}$ now reduce to 
$\int 4I\sqrt{8\pi p}P_m dz$.  Later we find it useful to change the
independent variable from $z$ to $P_m$.  We therefore write $dz =
dZ/dP_m dP_m$ and on changing the limits appropriately we obtain
\begin{displaymath}
\int^F_0 4I\sqrt{8\pi p(Z)}\left (-dZ/dP_m \right ) P_m dP_m\ .
\end{displaymath}
We now introduce $\Pi(Z) = \int^{Z}_0 4I\sqrt{8\pi p(z)}dz$ and after
an integration by parts in which $Z$ and $\Pi$ vanish at one or other
end point we have just $\int^F_0 \Pi(Z)dP_m$ in place  of the first
two terms of $8\pi \bar{W}$.

Using $Z(P_m)$ as our variable in the remaining terms we find that we
do not have to vary both the function $P_m(z)$ and its inverse
function $Z(P_m)$, but $\bar{W}$ still contains both $Z(P_m)$ and
$Z(P)$.  They are the same function (albeit of a different variable).
The third term of $8\pi \bar{W}$ in (\ref{eq38}) simplifies greatly
when integrated by parts
\begin{eqnarray*}
\int^F_0 \left [\int^{P_m}_0 \left ( \Phi/Z \right ) dP \right ]^2 
\left (-dZ/dP_m \right ) dP_m = \\ = 2 
\int^F_0 \Phi(P_m)\int^{P_m}_0 \Phi/Z dP dP_m\ .
\end{eqnarray*}
Again, we used the fact that $Z(F)$ is zero in the boundary terms.  We
reverse the order of integration using $\int^F_0\int^{P_m}_0 \left
  (\ldots \right ) dPdP_m = \int^F_0\int^F_P \left (\ldots \right )
dP_m dP $ to obtain
\begin{displaymath}
2\int^F_0 \Phi/Z \int^F_p \Phi (P_m)dP_m dP\ .
\end{displaymath}
Finally we exchange the dummy variables $P_m$ and $P$ to write the
third term of $\bar{W}$ in terms of $Z(P_m)$
\begin{equation}
8\pi \bar{W}_\phi = 2 \int^F_0 \Phi(P_m)\left [Z(P_m)\right
]^{-1}\int^F_{P_m}\Phi (P)dPdP_m \ .\label{eq39}
\end{equation}
Inserting all these simplifications into $8\pi \bar{W}$ and using
$\zeta(P_m) = Z^{-1}$ as our variable we find 
\begin{eqnarray}
8\pi \bar{W}= \int^F_0 \Pi d P_m + \nonumber \\
+ \left [J^2/(2\pi)\right ]\int^F_0 
\left [\Phi \zeta \int^F_{P_m} \Phi dP \right ]
dP_m + \nonumber \\
+ \int \left (F-P_m \right )^2 
\left (2\pi \right )^{-1}
\left (d\zeta/dP_m \right ) dP_m\ , \label{eq40}
\end{eqnarray}
where $\Pi = \Pi(Z)$ may equally be considered as a function of
$\zeta$. 
\subsection{Solution of the Variational Equations}\label{sec3.3}
Varying $\zeta(P_m)$ and demanding that $\bar{W}$ be a minimum is now
easy since  the second and third terms are linear in $\zeta$ and
$\delta \Pi/ \delta \zeta = -4IZ^2\sqrt{8\pi p(Z)}$.  Thus the
variational equation for $Z(P_m)$ reads 
\begin{equation}
4IZ^2 \sqrt{8\pi p(Z)} = (2\pi)^{-1} J^2 \Phi (P_m)
\int^F_{P_m} \Phi dP + \pi^{-1} (F-P_m)\ . \label{eq41}
\end{equation}

\begin{figure}
\begin{center}
\includegraphics[scale=0.75]{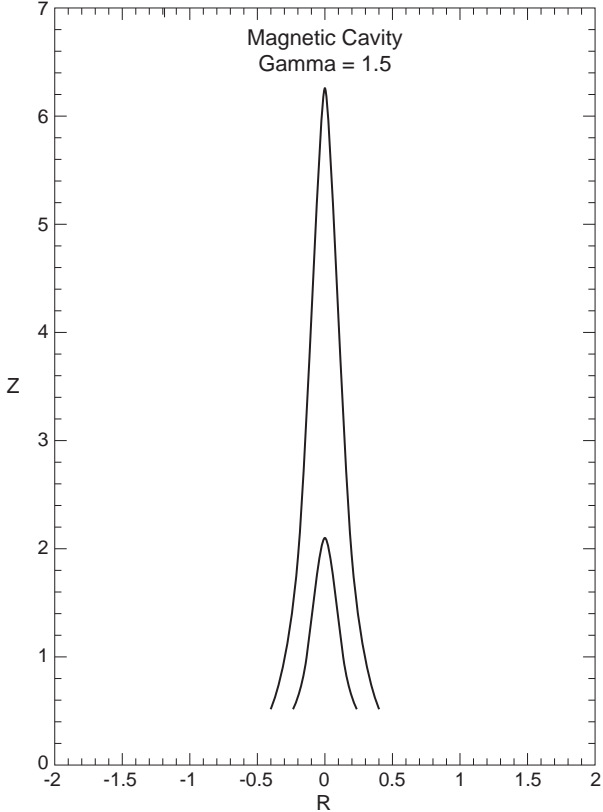}
\caption{Profiles of the Magnetic cavity in a constant pressure
  environment.  The differential rotation of the feet of the field
  lines labelled $P$ is $\Omega(P)\propto(P_0+P)^{-\gamma}$ with $P_0$
  small and $\gamma = 1.5$.  The profile of the cavity is given at two
  times.   }   
\end{center}
\end{figure}

\noindent
Thus, the function $Z(P_m)$ is given by 
\begin{eqnarray}
Z\left [p(Z)\right ]^{{\scriptscriptstyle{{1\over 4}}}}=\nonumber &\\
= (\pi/2)^{{\scriptscriptstyle{{1\over 4}}}} & \hspace{-10pt}
\left [{J\over (4\pi)}\right ] I^{-{\scriptscriptstyle{{1\over 2}}}} 
\left [
\begin{array}{l} 
\Omega(P_m) \int^F_{P_m}\Omega (P)dPt^2 +\\
+ \pi^{-1} \left (F - P\right )
\end{array} 
\right ]^{{\scriptscriptstyle{{1\over 2}}}} \ . \label{eq42}
\end{eqnarray}
The final $F-P$ term arises from our rough estimate of the $B_R$ term; 
unlike the $t^2$ term next door, it does not grow with time, so after
a few turns it is negligible as we expected.  Neglecting it we define
\begin{equation}
\bar{\Omega}(P_m)=
\left [\Omega (P_m)P^{-1}_m 
\int^F_{P_m}\Omega(P)dP \right ]^{{\scriptscriptstyle{{1\over 2}}}}  \
,  \label{eq43}
\end{equation}
and we obtain the very pleasing result
\begin{equation}
Z\left [p(Z)\right ]^{{\scriptscriptstyle{{1\over 4}}}}
 = C_1 P^{{\scriptscriptstyle{{1\over
        2}}}}_m  \bar{\Omega}(P_m)t \ , \label{eq44}
\end{equation}
where 
\begin{displaymath}
C_1 = (\pi/2 )^{{\scriptscriptstyle{{1\over 4}}}}\left [J/(4\pi)\right ]
I^{-{\scriptscriptstyle{{1\over 2}}} } \ .
\end{displaymath}
At given $P_m$ (\ref{eq44}) gives $Z$ as a function of $t$.  Knowing
that function we know how $p^{-{\scriptscriptstyle{{1\over 4}}}}$  
behaves with $t$.  That then
shows us how $Z/t$ behaves with $t$; (\ref{eq36}) shows us that, at
constant $P_m$, $R_m$ also behaves as $p^{-{\scriptscriptstyle{{1\over 4}}}}$. 
Since $p(z)$ is a given function this serves to define $Z(P_m)$ or for
that matter $P_m(z)$.  Of course we still have to derive the values of
the dimensionless structure functions $I$ and $J$ that we gave
earlier, but these only affect the time--scale of the evolution.
Using (\ref{eq36}) to re-express $P^{{\scriptscriptstyle{{1\over
        2}}}}_m$  in terms of $R_m$
(\ref{eq44}) takes the pleasing form 
\begin{equation}
Z={\scriptscriptstyle{1\over 4}}(J/I)R_m\bar{\Omega}t \label{eq45}
\end{equation}
which is a more sophisticated version of our result (\ref{eq14}).
Notice that ({\ref{eq42}) holds {\bf at each value of } $P_m$ thus the
  collimation $Z(P_m)/R(P_m)$ grows linearly with time even when the
  pressure depends on height.

\begin{figure}
\begin{center}
\includegraphics[scale=.750]{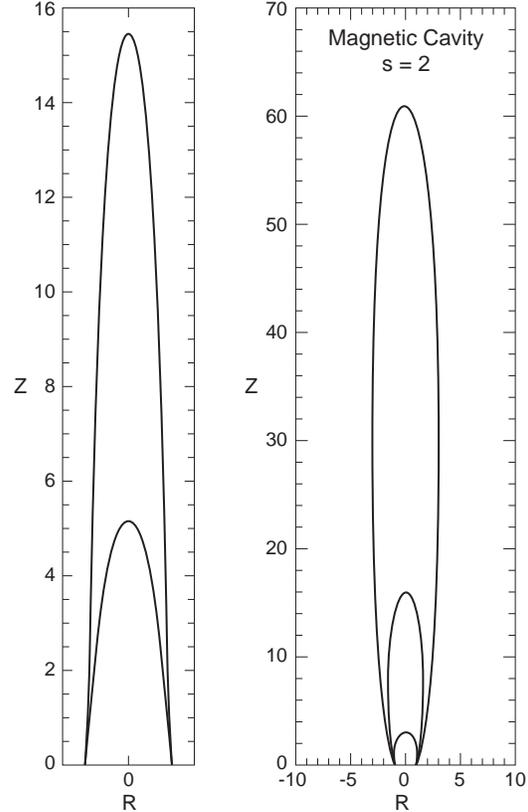}
\caption{The `central dipole' model gives a more rounded profile than
  the power law model.  4a is at constant pressure, 4b has pressure
  falling as $(1+z)^{-2}$ and is drawn for three times. }
\end{center}
\end{figure}

  The shapes of the magnetic cavities as functions of time follow
  directly by substituting the value of $P_m$ in terms of $R_m$ and
  $p(Z)$ from (\ref{eq36}) viz $P_m = \left [2\pi^3 p(Z)\right
  ]^{{\scriptscriptstyle{{1\over 2}}}}R^2_m/I$ into
  $\bar{\Omega}(P_m)$ in (\ref{eq44}).  After that (\ref{eq42})
  becomes an equation connecting the radius $R_m$ of the magnetic
  cavity to the height $Z$ at which the cavity has that radius.  We
  draw some examples of the evolution of the cavity shapes as
  functions of time in Figures 3 and 4.  When $p$ is independent of
  $z,\ R_m(P_m)$ is independent of time by (\ref{eq36}) so all the
  expansion takes place along
$z$ axis.  However when $p$ decreases with height $R_m(P_m)\propto
P^{\scriptscriptstyle{{1\over 2}}}_m \left [p(Z)\right
]^{-{\scriptscriptstyle{{1\over 4}}} }$    
and the last factor increases as
$z$ increases so it does  increase with time.  In spite of this $Z/R$
grows linearly with $t$ as seen in (\ref{eq45}).  A simple model will
illustrate this, we take $p(z)=p_0(1+z/a)^{-s}$.  We need $s\leq 4$ so
  that the field does not splay out as in paper I.  Equation
  (\ref{eq44}) for $Z$ now reads, writing
\linebreak  $V=(2\pi)^{{\scriptscriptstyle{{1\over 4}}} }\left
     [J/(2\pi)\right ]P^{{\scriptscriptstyle{{1\over 2}}}}_m \Omega
   (P_m) {Z \over (1+Z/a)^{s/4}} = 
  Vt$.  Taking $s=2$ as a simple example
\begin{displaymath}
Z=Vt \left [{Vt \over a} + \sqrt{1+V^2t^2/a^2}\right ]\ . 
\end{displaymath}
Thus, when the pressure falls asymptotically like $z^{-2},\ Z(t)$
starts out at constant velocity $V$ but then accelerates and
asymptotically it behaves as $2V^2t^2/a$ of course ram pressure and
inertia which are missing from our model may reduce this acceleration
and ram-pressure could even reverse it, nevertheless, without such our
model predicts acceleration when the pressure falls with height.  The
radius corresponding to $P_m$ is now given by 
\begin{displaymath}
R_m \propto P^{{\scriptscriptstyle{{1\over 2}}}}_m  
\left [{Vt \over a}+\sqrt{1+V^2t^2/a^2}\right ]\
. 
\end{displaymath}
For small $Vt/a$ this is proportional to $1+Vt/a$ but for large $Vt/a$
this factor changes to $2Vt/a$  so asymptotically $R_m$ is
proportional to time when $Z$ is proportional to $t^2$.  These results
are dependent on our choice of $s=2$.  For general $s$, $Z$
asymptotes to $Vt\left [Vt/a \right ]^{s/(4-s)}$ while $R_m$
asymptotes to $[Vt/a]]^{s/(4-s)}$ .  Both these expressions demonstrate
the very rapid expansion as $n$ approaches 1 i.e., when $p$ falls
almost as fast as $z^{-4}$ and indeed our former considerations were
limited to pressure that fell less fast than $r^{-4}$ in spherical
coordinates.  The great extent of radio jets might however be an
indication that we should be considering not merely enclosed fields
but those configurations that ``reach out to infinity''
We shall consider these again in another paper, but we remark here
that if a field carrying the poloidal flux $F$ reaches out to a region
beyond which the pressure behaves as $H r^{-4}$ then the field will
escape to infinity with $B^2_r = 8\pi H r^{-4}$ with an opening angle
$theta$ given by $2\pi \left [1-\cos (\theta/2)\right ] = F/\sqrt{8\pi
  H}$.  For small $F/\sqrt{H}$ such opening angles can be small
\begin{displaymath}
\theta = \left [2F^2/(\pi^3H)\right ]^{{\scriptscriptstyle{{1\over 4}}}}
\end{displaymath}
and in such configurations the field will be close to radial from the
source.  It is possible that we should be considering the large scale
radio jets as examples of this phenomenon rather than taking the
pressure to fall off less strongly than $r^{-4}$, but currently I feel
the other case to be more general.  
\subsection{Estimation of the Twist Function $\beta(P)$.}\label{sec3.4}
In the discussion following equation (\ref{eq30}) we showed that the
concentration of twist toward the top of each field--line behaved as
$(1-f)^{-{\scriptscriptstyle{{1\over 2}}}}$.  If we substitute this
behaviour for $d{\rm ln}(\lambda_i/\lambda_0)d{\rm ln}f $ in equation
(\ref{eq28}) and integrate along each line we find
\begin{displaymath}
\Phi(P)\propto {\scriptscriptstyle{1\over 2}}P^{-1}\beta (P)\int^F_P
\left [1-(P/P_m)\right ]^{-{\scriptscriptstyle{{1\over 2}}}} (-dZ/dP_m)dP_m\ .
\end{displaymath}
With $Z(P_m)$ now known we can deduce the form of $\beta(p)$
\begin{equation}
\beta(P)\propto 2P\Phi(P)\left /
\int^F_P 
{P^{{\scriptscriptstyle{{1\over 2}}}}_m (-dZ/dP_m)dP_m
\over (P_m - P)^{{\scriptscriptstyle{{1\over 2}}}}}\right. \label{eq46}
\end{equation}
We can not go further without specific models for $\Omega(P)$ and
$p(z)$. 
\subsection{Displaced Power Law Models}\label{sec3.5}
$\Omega(P)$ has to be zero at $P=F$ and it must fall as $P$ increases
from $0$ to $F$.  If $\Omega$ is finite at $P=0$ then the simplest
`power law' model is the doubly displaced one 
\begin{equation}
\Omega(P)= \Omega_0^{\ } P^\gamma_0 \left [
(P_0+ P)^{-\gamma} - 
(P_0 + F)^{-\gamma}\right ] \ . \label{eq47}
\end{equation}
We calculate in the regime $P_0 \ll F,\ F-P \gg P_0$ in which
circumstance the $(P_0+F)^{-\gamma}$ 
term is negligible.
\begin{eqnarray}
\Omega(P_m) \int^F_{P_m}\Omega(P)dP = 
{\Omega^2_0 P^{2\gamma}_0 \over 
  (\gamma -1)} 
\varpi^{-\gamma} 
\left [\varpi^{-(\gamma -1)} - \right.\nonumber\\
\left. - (P_0 +F)^{-(\gamma-1)}\right ]\ , \label{eq48}
\end{eqnarray}
where $\varpi = P_0+P_m$.   From (\ref{eq44}) the above expression is
proportional to $Z[p(Z)]^{{\scriptscriptstyle{{1\over 4}}}}$.   
Two cases give power laws; $\gamma >
1$ gives $Zp^{{\scriptscriptstyle{{1\over 4}}}}\propto \varpi^{ -(\gamma -
  {\scriptscriptstyle{{1\over 2}}}) }t$, while $0<\gamma
<1$ gives $Zp^{{\scriptscriptstyle{{1\over 4}}}}
\propto \varpi^{-\gamma/2}$.  We take our former
displaced power law for the pressure with $s<4$
\begin{displaymath}
p=p_0\left [(z/a)+1\right ]^{-s}\ .
\end{displaymath}
Notice that, for $z<a/s$, the pressure is almost constant so that regime
is equivalent to taking $s=0$ but, when $z$ is large, $p\propto z^{-s}$
so $Z\left [p(Z)\right ]^{{\scriptscriptstyle{{1\over 4}}}}  
\propto Z^{(4-s)/4}$.  Hence we have the
power laws 
\begin{equation}
Z\propto \varpi^{-\Gamma}t^S\ , \label{eq49}
\end{equation}
where
\begin{displaymath}
\Gamma =
\left \{
\begin{array}{lll}
(\gamma -{\scriptscriptstyle{1\over 2}})S & {\rm for}\ \gamma >1 &
{\rm and}\ S=1/[(s/4)]\ . \\
{\scriptscriptstyle{1\over 2}}\gamma S & {\rm for}\ \gamma <1 & \\
\end{array}
\right. 
\end{displaymath}
Rewriting (\ref{eq45}) in terms of $\varpi$ setting $\varpi_\ast = P_0
+ P$
\begin{displaymath}
\beta \propto 2P\Omega(P)t\left /\int^{F+P_0}_{\varpi_\ast} 
{(\varpi - P_0)^{{\scriptscriptstyle{{1\over 2}}}} 
(-dZ/d\varpi) \over (\varpi -\varpi_\ast)^{{\scriptscriptstyle{{1\over
        2}}}}} \right. d\varpi\ . 
\end{displaymath}
Now write $x=\varpi/\varpi_\ast$ and recall $\Omega(P)\propto
\varpi_\ast^{-\gamma}$. 
\begin{equation}
\beta \propto P\varpi_\ast^{\Gamma -\gamma}M t^{1-S}\ , \label{eq50}
\end{equation}
where
\begin{displaymath}
M=\Gamma \int^\infty_1 
{\left [x-(P_0/\varpi_\ast)\right ]^{{\scriptscriptstyle{{1\over 2}}}}
x^{-(\Gamma + 1)}dx \over (x -1)^{{\scriptscriptstyle{{1\over 2}}}}}
\end{displaymath}
and we have assumed $\left [(F+P_0)/(P+P_0)\right ]\gg 1$ to allow the
upper limit to tend to infinity. 

Notice that, for $P \gg P_0$, $M$ is independent of $P$ and indeed $M=
\pi^{2}\Gamma ! /(\Gamma- {\scriptscriptstyle{{1 \over 2}}})!$ which
is $3\pi/4$ for $\Gamma = {\scriptscriptstyle{{3\over 2}}}$ and
tends to $1$ as $\Gamma \rightarrow 1$; for $P=0$, $M=1$.

For the constant pressure case $s=0$, $S=1$, and when $\gamma >1$ 
\begin{equation}
\beta \propto P/(P_0+P)^{{\scriptscriptstyle{{1\over 2}}}} \ , \label{eq51}
\end{equation}
so when $P_0$ is small compared with $P,$ $\beta \propto
P^{{\scriptscriptstyle{{1\over 2}}} }$ 
\begin{equation}
{\rm In\ general}\ \Gamma - \gamma =
\left \{
\begin{array}{ll}
\gamma(S-1)-{\scriptscriptstyle{1\over 2}}S & \gamma >1 \\
{\scriptscriptstyle{1\over 2}}\gamma(S-2)   & 0<\gamma<1
\end{array}\ .
\label{eq52}
\right.
\end{equation}
\begin{displaymath}
{\rm For}\ s=2,\ S=2\ {\rm so}\ \Gamma - \gamma =
\left \{
\begin{array}{ll}
\gamma - 1 & \gamma >0 \\
0 & 0<\gamma <1
\end{array}
\right.
\end{displaymath}

In paper IV we shall solve for the fields in such power law cases.
Particular interest
centres around cases in which $\beta$ is proportional to
$P^{{\scriptscriptstyle{{1\over 2}}} }$ or $P$ which give rise to
linear partial differential equations for $P$ c.f. equation
(\ref{eq24}).  The former occurs when pressure is constant i.e., $S=1$
when $\gamma >1$ wherever $P_0$ is small compared with $P$.  This has
solution $f=-e\lambda {\rm ln}\lambda$.  The latter $(\beta \propto
P)$ occurs when $S=2\gamma/(2\gamma -1)$ and $\gamma >1$ i.e., when
$s=2/\gamma$ and with $S=s=2$ with $\gamma <1$.  Although the other
power laws give non--linear equations they too can be solved except
near the top of the tower.  Again we leave the discussion to paper IV.
Rotation in a gravity--dominated disk cannot fall off more rapidly
than $\Omega_d \propto R^{-{\scriptscriptstyle{{3\over 2 }}}}$ while
if the $B_z$ varies as $R^{-\eta}$ on the disk then $P\propto
R^{2-\eta}$.  Thus $\Omega_d(P)\propto P^{-3/(4-2\eta)}$.  The
requirement of a finite field on axis suggests that near there $\eta =
0$ and $\Omega_d \propto P^{-{\scriptscriptstyle{{3\over 4}}} }$ i.e.,
$\eta = {\scriptscriptstyle{{3\over 4}}}$ but if the field further out
falls off as $R^{-{\scriptscriptstyle{{3\over 2}}}}$ then $\eta =
{\scriptscriptstyle{{3\over 2}}}$ and $\gamma = 3$ so $\gamma$ is
likely to be $>1$ because $B_z$ is likely to fall faster than
$R^{-{\scriptscriptstyle{{1\over 2}}}}$
\subsection{Models with a Central Dipole}\label{sec3.6}
If a central object of radius $R_s$ rotates uniformly at angular
velocity $\Omega_s$
and carries a dipole of strength
$D$
then initially the field at radius
$R$ 
 on the disk is 
$B_z = D/R^3$ for
$R>R_s$.  If the accretion disk is dominated by the central object then in
centrifugal force gravity balance
\begin{displaymath}
\Omega_d = \left (GM_s/R^3\right)^{{\scriptscriptstyle{{1\over 2}}}}
\end{displaymath}
Now before it is twisted the flux function
$P$
of the dipole is
$P=2\pi D r^{-1}\sin^2\theta$ 
so on the disk
$P=2\pi DR^{-1}$
for
$R>R_s$.  
Now suppose 
$\Omega_s <(GM_s/R_s^3)^{\scriptscriptstyle{{1\over 2}}}{}$ 
then there will be a co--rotation point in the disk at some point
$R=R_c=[\Omega^2_s/(GM_s)]^{{\scriptscriptstyle{{1\over 3}}}}$
where
$P=P_c = 2\pi DR^{-1}_c$
.  

Then for 
$P>P_c$
we shall have a differential winding of the flux feet with
$\Omega_s - \Omega_d$ 
negative while for 
$P>P_c$
we shall have a differential winding with 
$\Omega_s - \Omega_d$
positive.

So
\begin{equation}
\Omega (P) = \Omega_s - (GM_s)^{{\scriptscriptstyle{{1\over 2}}}}\left
  [{P\over (2\pi D)}\right ]^{{\scriptscriptstyle{{3\over 2}}}} 
= \Omega_s \left [1-
\left ({P \over P_c}\right )^{{\scriptscriptstyle{{3\over 2}}}}\right
]\ . \label{eq53}  
\end{equation}
Assuming that all the flux is wound up there will no doubt be
continual reconnection close to
$P=P_c$
where the toroidal flux runs in opposite directions.  It then becomes
of interest to ask whether a greater poloidal flux crosses the equator
in the range
$R_s$
to
$R_c$
or in the range
$R_0$
to
$\infty$
.  Since
$P$
falls like
$R^{-1}$
outside the star half the flux is within
$R_{q/w}=2R_s$
so if the star rotates at less than 
$\sqrt{8}(GM_s/R^3_s)^{{\scriptscriptstyle{{1\over 2}}}}$
then the majority of the poloidal flux is dragged forward by the disk
while if the star rotates at closer to breakup speed the disk will be
dragging the majority of the flux backward.  Of course the central
object's field may be so strong that it enforces co--rotation as the
inner disk without being able to do so further out.  There is yet
another interesting radius in the problem.  if we think of the
oppositely wound fields then at what radius is the rate of generation
of toroidal flux within balanced by the rate of generation of net
toroidal flux of the other sign on the outside?  A little thought
shows this can only occur when $R_c = ({\scriptscriptstyle{5 \over
    2}})^{{\scriptscriptstyle{2\over 3}}} R_s = 1.84 R_s$.  On
integration (\ref{eq53}) gives, using (\ref{eq42}) with $P_c = F$ 
\begin{displaymath}
Z^2\sqrt{p(Z)} = \left ({1\over 4\pi}\right )^2 \left ({\pi \over
    2}\right )^{\scriptscriptstyle{1\over 2}}J^2I^{-1}Fy\bar{\Omega}^2\ ,
\end{displaymath}
where $y=P_m/F$
\begin{displaymath}
\bar{\Omega}= \Omega_s \left \{ y^{-1}\left
    [1-y-{\scriptscriptstyle{2\over 5}}\left
    (1-y^{{\scriptscriptstyle{5\over 2}}}\right )\right ]\ 
\left [1-y^{{\scriptscriptstyle{5\over 2}}}\right ] \right
    \}^{\scriptscriptstyle{1\over 2}} 
\end{displaymath}
\section{An Improvement in Accuracy}\label{sec4}
We showed earlier that the twist of the field line labelled $P$ is
given by  
\begin{displaymath}
d\Phi/dz = {\scriptscriptstyle{1\over 2}}\beta (P) P^{-1} d{\rm
  ln}\left ({\lambda_i/\lambda_0}\right )/d{\rm ln}f \ , 
\end{displaymath}
and that this last factor had an integrable infinity that behaved as
 $(1-f)^{-{\scriptscriptstyle{1\over 2}}} =
[1-(P/P_m)]^{-{\scriptscriptstyle{1\over 2}}} $.  Here we show how to
  take some account of such a distribution of twist with height.  In
  the total twist we shall have the factor
\begin{eqnarray*}
\int^Z_0[1-(P/P_m)]^{-{\scriptscriptstyle{1\over 2}}} dz = Z + \hspace{10pt} 
&\\ 
+
  \int^{{\rm ln}F}_{{\rm ln}P}\left \{\left [ 1-(P/P_m)\right
  ]^{-{\scriptscriptstyle{1\over 2}}}-1 \right \} &\hspace{-10pt} \left (-dZ/d{\rm
  ln}P_m \right )d{\rm ln}P_m\ .
\end{eqnarray*}
The quantity within   $\left \{ \ \ \right \}$ is small when $(P/P_m)$
is small but becomes large close to $P_m =P$.  Near there
$-dZ(P_m)/d{\rm ln}P_m$ can be approximated as constant at the value
$-dZ/d{\rm ln}P$.  Since most of the contribution to the integral
comes from near there we may take $-dZ/d{\rm ln}P_m$ constant at that
value and move it outside the integration which then becomes writing
$x=P_m/P $ and $1-x^{-1} = y^2$ 
\begin{eqnarray*}
{\cal{L}}(P) & = &\int^{F/P}_0 \left [\left (1-{\scriptscriptstyle{1\over
        x}} \right )^{-{\scriptscriptstyle{1 \over 2}}}- 1 \right ]
        x^{-1}dx =\hspace{50pt} \\
& = & \int^{\sqrt{1-P/F}}_0 (1+y)^{-1} 2dy 
= 2 {\rm 
        ln}\left [1+\sqrt{1-P/F}\right ] \   . 
\end{eqnarray*}
The small region where $P$ is close to $F$ inhabits very little of the
volume as these lines of force only just emerge from the disk before
returning to it and, for $P$ small, this integral is close to ${\rm ln}
4 = 1.38$.  We may take account of the increased weight near the top
of each field line by taking the distribution of twist with height to
be the sum of a uniform distribution and a 
$\delta$ function just below the top.  Thus
\begin{equation}
g(P,z)dz = \left [ {1+ \delta(z-Z_{-})(-dZ/d{\rm ln}P){\cal L}(P)}\right
]N_1^{-1}Z^{-1}dz \ , \label{eq54}
\end{equation}
where $Z_{-}$ is $Z(P)$ minus a very small quantity so that integrating
up to $Z(P)$ integrates over the $\delta$ function.  $N_1$ is the
normalising factor that ensures that $g$ integrates to 1.
\begin{displaymath}
N_1=\left [1+(-d{\rm ln}Z/d{\rm ln}P){\cal L}(P)\right ]\ .
\end{displaymath}
Now $Z$ decreases as $P$ increases so $-d{\rm ln}Z/d{\rm ln}P$ is
positive and will show no great variation. Since ${\cal L}(P)$ is
around unity for much of the range and $N_1$ is bounded below by 1 we
shall simplify by replacing $-d{\rm ln}Z/d{\rm ln}P$ by $-d{\rm
  ln}Z_0/d{\rm ln}P$ in the expression for $N_1$, where $Z_0$ is the
approximate solution given by (\ref{eq44}).  We insert the above $g$
into (\ref{eq32}) and find in place of (\ref{eq33})
\begin{displaymath}
\langle B^2_\phi \rangle = \left [J/(2\pi)\right ]^2 
\left [
\begin{array}{l}
\Phi P_m {\cal L}(P_m)N_1^{-1}Z^{-1}+ \\
\int^{P_m}_0 \Phi/(N_1Z)dP 
\end{array}
\right ]^2 R^{-2}_m \ . 
\end{displaymath}
Writing $\zeta$ for $Z^{-1}$ our expression for $8\pi W_\phi$ becomes 
\begin{displaymath}
(4\pi)^{-1}J^2 \int^F_0
\left [
\begin{array}{l}
\Phi^2 P^2_m {\cal L}^2 N_1^{-2}{d\zeta\over d P_m} + \\
+ 2\Phi P_m {\cal L} N_1^{-1}{d{\rm ln}\zeta\over dP_m} 
\int^{P_m}_0 \Phi \zeta
N_1^{-1}dP + \\ + 2 \zeta \Phi N_1^{-1} \int^F_{P_m}\Phi {N_1^{-1}}dP 
\end{array}
\right ]
dP_m \ . 
\end{displaymath}
The first and last terms are linear in $\zeta$ so their variation is
simple.  The central term while homogeneous of degree 1 in $\zeta$ is
not linear because of the $d{\rm ln}\zeta/dP_m$ term.  Varying this 
central term inside the integral and writing $\Psi = d/dP(\Phi P {\cal
  L}/N_1)$ we find 
\begin{displaymath}
\int^F_0 - 2\delta \zeta 
\left \{
\begin{array}{l}
\Phi^2N_1^{-2}P_m {\cal L} + \\
+ Z \Psi(P_m)\int^{P_m}_0 \Phi/(N_1Z)dP -\\
- \Phi N_1^{-1} \int^F_{P_m} {\rm ln}\left [ Z(P)/Z(P_m)\right ]
\Psi(P)dP 
\end{array}
\right \}
dP_m \ .
\end{displaymath}
The coefficient of $\delta \zeta$ is homogeneous of degree zero in $Z$
but causes  significant difficulties when one tries to solve the
equations for the minimising $Z(P_m)$.  We surmount these difficulties
by evaluating these terms with our zero order solution $Z=Z_0(P_m)$ as
given by (\ref{eq44}).  If greater accuracy is required we can then
solve and put back the solution iteratively.

The variation of $8\pi W_\phi$ as given above leads us to the
following equation for $\delta(8\pi W)$ in place of (\ref{eq41})
\begin{eqnarray}
\begin{array}{l}
4I\sqrt{8\pi Z^4p(Z)} = \\ 
= (2\pi)^{-1} J 
\end{array}
\left \{ 
\begin{array}{l}
N_1^{-1}\Phi(P_m)\int^F_{P_m}{\Phi\over N_1} {dP} -  \\
- \Phi N_1^{-1}{\cal L}P_m \Psi (P_m)- \\
- \Phi^2N_1^{-2}{\cal L} P_m - \\
-Z_0 \Psi(P_m)\int^{P_m}_0 {\Phi dP\over (N_1Z_0)} +\\
+ ({\Phi\over N_1} )
\int^F_{P_m}\Psi(P)\times \\
\times {\rm ln} \left [{Z_0(P)\over Z_0(P_m)}\right ]dP \ .
\end{array}
\right \} \label{eq55} 
\end{eqnarray}

All terms on the right are known so this gives us a refined solution
for $Z(P_m)$.  Every variable outside an integral is a function of
$P_m$ and, except where stated explicitly, variables inside an
integral are functions of $P$.  While the above expression may be more
accurate it provides far less understanding than our former solution
(\ref{eq44}).  As our aim should be {\bf insight} rather than
accuracy we have used (\ref{eq44}) in preference to (\ref{eq55}).  

A method of improving on the solution (\ref{eq55}) so that the
variable twist with height is properly allowed for is given in the
Appendix.  
\section{Conclusions}
A sequence of static models can be more illuminating than detailed
dynamical simulations.  The statics should be understood before the
dynamics is attempted.  The continued winding of the magnetic field of
an accretion disk will build up tall towers which make magnetic
cavities, towering bubbles of magnetic field in the surrounding
medium.  Such a conclusion does not depend on axial symmetry.

Non--axysymmetric behaviour is observed in the pretty plasma
experiments of \citet{hsu02}.

When that symmetry is imposed we can calculate the tall tower shapes
of the magnetic cavities for any prescribed winding angles $\Phi(P)
=\Omega(P)t$ and for any prescribed pressure distribution $p(z)$.
Examples of these towers are shown in figures 3 and 4.  The heights of
these towers grow at a velocity closely related to the maximum
circular velocity in the accretion disk.  Our primary results are
encapsulated in equations (\ref{eq44}) and (\ref{eq45}).  

As stellar--mass black--holes form, the winding of the magnetic field
of the collapsing core may cause jets to emerge from supernovae as
first predicted by \citet{leb70}.  Such ideas may have application to
some $\gamma$--ray bursts, to the micro--quasars \citep{mir99} and to
SS433 \citep{mar84}.  The possibility that the elongated hour--glass
planetary nebulae, some of the most delicate objects in the sky, may
be magnetic towers arising from the accretion disks of their central
binaries is particularly appealing.  

\section{Acknowledgements}
I thank Peter Goldreich and Houshang Ardavan for their encouragement
and criticism.  Mark Morris suggested that magnetic tower solutions
might be applied to accretion disks  of the binary stars inhabiting
the elongated hour--glass planetary nebulae.

\section*{References}
\begin{itemize}
\bibitem[\protect\citeauthoryear{Aly}{1984, 1995}]{aly8495}
Aly, J.J., 1984, ApJ, 283, 349
\bibitem[\protect\citeauthoryear{Aly}{1995}]{aly95}
Aly, J.J., 1995, ApJ, 439, L63
\bibitem[\protect\citeauthoryear{Appl \& Camenzind}{1993}]{app93}
Appl, S. \& Camenzind, M., 1993, A\&A, 270, 71
\bibitem[\protect\citeauthoryear{Baade}{1956}]{baa56}
Baade, W., 1956, ApJ, 123, 550
\bibitem[\protect\citeauthoryear{Barnes \& Sturrock}{1972}]{bar72}
Barnes, C.W. \& Sturrock, P.A., 1972, ApJ, 174, 659
\bibitem[\protect\citeauthoryear{Begelmann et al.}{1984}]{beg84}
Begelman, M.C., Blandford, R.D. \& Rees, M.J., 1984, Revs.Mod.Phys,
56, 255
\bibitem[\protect\citeauthoryear{Bell \& Lucek}{1995, 1996}]{bel9596}
Bell, A.R. \& Lucek, S.G., 1995, MNRAS, 277, 1327
\bibitem[\protect\citeauthoryear{Bell \& Lucek}{1996}]{bel96}
Bell, A.R. \& Lucek, S.G., 1996, MNRAS, 290, 327
\bibitem[\protect\citeauthoryear{Blandford \& Payne}{1982}]{bla82}
Blandford, R.D. \& Payne, D.G., 1982, MNRAS, 199, 883
\bibitem[\protect\citeauthoryear{Blandford \& Rees}{1974}]{bla74}
Blandford, R.D. \& Rees, M.J., 1974, MNRAS, 169, 395
\bibitem[\protect\citeauthoryear{Blandford \& Znajek}{1977}]{bla77}
Blandford, R.D. \& Znajek, R.L., 1977, MNRAS, 179, 433
\bibitem[\protect\citeauthoryear{Contopoulos}{1995a}]{con95a}
Contopoulos, J., 1995a, ApJ, 446, 67
\bibitem[\protect\citeauthoryear{Contopoulos}{1995b}]{con95b}
Contopoulos, J., 1995b, ApJ, 450, 616
\bibitem[\protect\citeauthoryear{Curtis}{1918}]{cur18}
Curtis, H.D., 1918, Pub. Lick Observatory, 13, 31
\bibitem[\protect\citeauthoryear{Eddington}{1926}]{ed26}
Eddington, A.S., 1926, The Internal Constitution of the Stars, p109,
Cambridge Univ. Press
\bibitem[\protect\citeauthoryear{Ekers \& Lynden-Bell}{1971}]{eke71}
Ekers, R. \& Lynden-Bell, D., 1971, Astrophysical Lett., 9, 189
\bibitem[\protect\citeauthoryear{Heyvaerts \& Norman}{1989}]{hey89}
Heyvaerts, J. \& Norman, C.A., 1989, ApJ, 347, 1055
\bibitem[\protect\citeauthoryear{Hoyle \& Ireland}{1960}]{hoy60}
Hoyle, F. \& Ireland, J.G., 1960, MNRAS, 120, 173
\bibitem[\protect\citeauthoryear{Hsu \& Bellan}{2002}]{hsu02}
Hsu, S.C. \& Bellan, P.M., 2002, MNRAS, 334, 257
\bibitem[\protect\citeauthoryear{Krasnopolskij et al.}{1999}]{kra99}
Krasnopolskij, R., Li, Z-Y., Blandford, R.D., 1999, ApJ, 526, 631
\bibitem[\protect\citeauthoryear{Le Blanc \& Wilson}{1970}]{leb70}
Le Blanc, J.M. \& Wilson, J.R., 1970, ApJ, 161, 541
\bibitem[\protect\citeauthoryear{Li et al.}{2001}]{li01}
Li, H., Lovelace, R.V.E., Finn, J.H. \& Colegate, S.A., 2001, ApJ,
5611, 915
\bibitem[\protect\citeauthoryear{Lovelace}{1976}]{lov76}
Lovelace, R.V.E., 1976, Nature, 262, 649
\bibitem[\protect\citeauthoryear{Lovelace et al.}{1991}]{lov91}
Lovelace, R.V.E., Beck, H.L. \& Contopoulos, J., 1991, ApJ, 379, 696
\bibitem[\protect\citeauthoryear{Lovelace et al.}{2002}]{lov02}
Lovelace, R.V.E., Li, H., Ustyugova, G.V. \& Romanova, M.M., 2002,
ApJ, 572, 445
\bibitem[\protect\citeauthoryear{Lynden-Bell}{1969, 1971}]{dlb6971}
Lynden-Bell, D., 1969, Nature, 223, 690
\bibitem[\protect\citeauthoryear{Lynden-Bell}{1971}]{dlb71}
Lynden-Bell, D., 1971, MNRAS, 155, 199
\bibitem[\protect\citeauthoryear{Lynden-Bell}{1996}]{dlb96}
Lynden-Bell, D., 1996, MNRAS, 279, 389, {\bf Paper II}
\bibitem[\protect\citeauthoryear{Lynden-Bell}{2001}]{dlb01}
Lynden-Bell, D., 2001, ASP Conference 249, 212, The Central kpc of
Starbursts and AGN, eds. Knapen, J.H, Beckman, J.E., Shlosman, I. \&
Mahoney, T.J. astro-ph/0203480 
\bibitem[\protect\citeauthoryear{Lynden-Bell}{in preparation}]{dlb02}
Lynden-Bell, D.,  {\bf Paper IV}
\bibitem[\protect\citeauthoryear{Lynden-Bell \& Boily}{1994 Paper I}]{dlb94}
Lynden-Bell, D. \& Boily, C., 1994, MNRAS, 267, 146, {\bf Paper I}
\bibitem[\protect\citeauthoryear{Lynden-Bell \& Pringle}{1974}]{dlb74}
Lynden-Bell, D. \& Pringle, J.E., 1974, MNRAS, 168, 603
\bibitem[\protect\citeauthoryear{Lynden-Bell \& Rees}{1972}]{dlb72}
Lynden-Bell, D. \& Rees, M.J., 1972, MNRAS, 152, 461
\bibitem[\protect\citeauthoryear{Mirabel \& Rodriguez}{1999}]{mir99}
Mirabel, I.F. \& Rodriguez, F.J., 1999, Ann.Rev. of A\&A, 37, 409
\bibitem[\protect\citeauthoryear{Margon}{1984}]{mar84}
Margon, B., 1984, Ann.Rev. of A\&A, 22, 507
\bibitem[\protect\citeauthoryear{Okamoto}{1999}]{oka99}
Okamoto, I., 1999, MNRAS, 307, 253
\bibitem[\protect\citeauthoryear{Ouyed et al.}{1997}]{ouy97}
Ouyed, R., Pudritz, R.E., \& Stone, J.E., 1997, Nature, 385, 409 
\bibitem[\protect\citeauthoryear{Ouyed \& Pudritz}{1999}]{ouy99}
Ouyed, R., \& Pudritz, R.E., 1999, MNRAS, 309, 233
\bibitem[\protect\citeauthoryear{Parker}{1965}]{par65}
Parker, E.N., 1965, ApJ, 142, 584
\bibitem[\protect\citeauthoryear{Pudritz \& Norman}{1987}]{pud87}
Pudritz, R.E. \& Norman, C.A., 1987, ApJ, 301, 571
\bibitem[\protect\citeauthoryear{Romanova et al.}{1998}]{rom98}
Romanova, M.M., Ustyugova, G.V., Koldara, A.V., Chechetkin, V.M. \&
Lovelace, R.V.E., 1998, ApJ, 500, 703
\bibitem[\protect\citeauthoryear{Rees}{1971}]{ree71}
Rees, M.J., 1971, Nature, 229, 312
\bibitem[\protect\citeauthoryear{Sakarai}{1985, 1987}]{sak8587}
Sakarai, R., 1985, A\&A, 152, 121
\bibitem[\protect\citeauthoryear{Sakarai}{1987}]{sak87}
Sakarai, R., 1987, PASJ, 39, 821
\bibitem[\protect\citeauthoryear{Scheuer}{1974}]{sch74}
Scheuer, P.A.G., 1974, MNRAS, 166, 513
\bibitem[\protect\citeauthoryear{Schott}{1912}]{sch12}
Schott, G.A., 1912, Electromagnetic Radiation \& Reactions Arising
from it, Cambridge Univ. Press
\bibitem[\protect\citeauthoryear{Shibata \& Uchida}{1985, 1986}]{shi8586}
Shibata, K. \& Uchida, Y., 1985, Pub.Ast.Soc.Japan, 37, 31
\bibitem[\protect\citeauthoryear{Shibata \& Uchida}{1986}]{shi86}
Shibata, K. \& Uchida, Y., 1986, Pub.Ast.Soc.Japan, 38, 631
\bibitem[\protect\citeauthoryear{Sturrock}{1991}]{stu91}
Sturrock, P.A., 1991, ApJ, 180, 655
\bibitem[\protect\citeauthoryear{Uzdensky et al.}{2002}]{uzab}
Uzdensky, D.A., Konigl, A. \& Litwin, C., 2002, (a) ApJ, 565, 1191, (b)
ApJ, 565, 1205 
\bibitem[\protect\citeauthoryear{Uzdensky}{2002}]{uz02}
Uzdensky, D.A., 2002, ApJ, 575, 432
\bibitem[\protect\citeauthoryear{Wheeler}{1971}]{whe71}
Wheeler, J.A., 1971, In Nuclei of Galaxies, p539, ed. O'Connell,
D.J.R., North Holland, Amsterdam
\end{itemize}
\appendix
\section*{Appendix}
\section{Perturbation Theory for the Variable Twist Problem}
We would like to give the field lines the correct distribution of
twist with height approximately
\begin{equation}
g(P,\ z) = \left [1-(P/P_m)\right ]^{-{\scriptscriptstyle{1\over 2}}}
dz /L(P)\ , \label{eqap1}
\end{equation}
where
\begin{eqnarray*}
L(P)& = & \int^Z_0\left [1-\left (P/P_m(z)\right )\right
]^{-{\scriptscriptstyle{1\over 2}}}
dz \ \ = \\
& = & \int^F_P\left [1-(P/P_m)\right ]^{-{\scriptscriptstyle{1\over
      2}}} (-dZ/P_m)dP_m \ ,
\end{eqnarray*}
putting this into (\ref{eq35}) gives in place of (\ref{eq37}) 
\begin{displaymath}
\langle B^2_\phi \rangle = J^2 
\left \{
\begin{array}{l}
\int^{P_m}_0 (2\pi)^{-1} \Phi(P) \times \\
\hspace{1cm}\times \left [1-({P\over P_m})\right
]^{-{\scriptscriptstyle{1\over 2}}} L^{-1}dP 
\end{array}
\right \}^2 \ \ R^{-2}_m \ , 
\end{displaymath}
which in turn gives an $8\pi W_\phi$ 
\begin{eqnarray}
8\pi W_\phi  
= (4\pi)^{-1}J^2 \times \nonumber \\ \times \int^F_0 
\left \{
\begin{array}{l}
\int^{P_m}_0 \Phi(P) [1- \left ({P\over P_m}\right )
]^{-{\scriptscriptstyle{1\over 2}}} L^{-1}dP 
\end{array}
\right \}^2 \times \nonumber \\
\times \left ({-dZ \over dP_m}\right ) dP_m \ .  \label{eqap2}
\end{eqnarray}
Here the function to be varied $Z(P_m)$ is deeply embedded in the
integrals that  define  $L(P_m)$ as well as occurring explicitly.
There is no simple way of solving the variational equations that
result, that is why we chose to consider the simpler forms of $g$
given in equations (\ref{eq31}) and (\ref{eq54}).  However the
variational principle (\ref{eq40}) based on (\ref{eq31}) is so simple
that a perturbation theory based on it can be developed.  We write $W$
for the energy based on (\ref{eqap2}) and $\bar{W}$ for the energy
based on (\ref{eq39}) so that
\begin{equation}
W[\zeta] = \bar{W}[\zeta] + (W_\phi - \bar{W}_\phi) \ . \label{eqap3}
\end{equation} 
Now $8\pi \bar{W}$ minimises at $\zeta_0$ and only the first $\Pi$
term is non--linear so we may write
\begin{eqnarray}
8\pi\bar{W}[\zeta]
= 8\pi \bar{W}[\zeta_0] + \nonumber \\ + \int^F_0
{\scriptscriptstyle{1\over 2}} \delta^2 \Pi/8\zeta^2 (\zeta
-\zeta_0)^2 + {{\scriptscriptstyle{1\over
      6}}}\delta^3\Pi/\delta\zeta^3 (\zeta -\zeta_0)^3 dP_m \
. \label{eqap4} 
\end{eqnarray}
Here $\zeta_0 = 1/Z_0$ which is the solution given by (\ref{eq44}).
We believe that $\zeta_1 = 1/Z_1$ where $Z_1$ is given by (\ref{eq55})
will be close to the minimum of $W$.  Our problem is to find the
correction of $\zeta_1$ which will give $\delta W/\delta\zeta =0$ more
accurately.  Evidently from (\ref{eqap3}) 
\begin{eqnarray*}
\delta W/\delta \zeta 
=  \delta \bar{W}/\delta\zeta + \delta(W_\phi -
\bar{W}_\phi)/\delta \zeta \ .
\end{eqnarray*}
Hence using (\ref{eqap4}) and evaluating $\delta(W_\phi-\bar{W}_\phi
)$ etc., at the approximate solution $\zeta_1$ we find 
\begin{eqnarray*}
\zeta - \zeta_1 = -\hspace{10cm} &\hspace{25cm} \\
\hspace{-35cm}- 
{ 
\left [8\pi\delta 
\left (W_\phi - \bar{W}_\phi \right )
/\delta\zeta_1 + \delta^2\Pi/\delta\zeta^2_0 
\left (\zeta_1 - \zeta_0 
\right )
+ {\scriptscriptstyle{1\over 2}} 
\delta^3\Pi/\delta \zeta_0^3 (\zeta_1 - \zeta_0 )^ 2
\right ]
\over
\delta^2\Pi/\delta\zeta^2_0 + 
\delta^3\Pi/\delta\zeta^3_0
(\zeta_1-\zeta_0)
}
\ ,  \hspace{1.8cm} & 
\end{eqnarray*} 
where we have written $\delta^2\Pi/\delta\zeta^2_0$ for
$\delta^2\Pi/\delta\zeta^2$ evaluated at $\zeta_0$ etc.  Now
$\delta\Pi/\delta\zeta = - 4IZ^2\sqrt{8\pi p(Z)}$ so
\begin{displaymath}
\delta^2\Pi/\delta\zeta^2 = 4IZ^2 \partial/\partial Z
\left [Z^2 \sqrt{8\pi p(Z)} \right ] 
\end{displaymath}
and 
\begin{displaymath}
\delta^3 \Pi/\delta\zeta^3 = - 4I Z^2 d/dZ \left [Z^2
  d/dZ(Z^2\sqrt{8\pi p(Z)})\right ] \ , 
\end{displaymath}
so all those quantities are quite simple functions of $Z$ and we have
to evaluate them at $Z=Z_0$ the solution (\ref{eq44}).  Thus all terms
on the right are known except
$8\pi\delta(W_\phi-\bar{W}_\phi)/\delta\zeta_1$ which we have to
evaluate at $\zeta_1$.  We refrain from giving the gory details of the
evaluation but 
\begin{displaymath}
{8\pi \delta(W_\phi -\bar{W}_\phi)\over \delta \zeta} = -
2\pi J^2 EZ\ .\ {d(EZ)\over dP_m} + {\delta Y\over \delta\zeta } -
2P_m\bar{\Omega}^2 \ , 
\end{displaymath}
where $E(P_m)=\int^{P_m}_0\left (\Phi/2\pi L\right )\left [1-(P/P_m)\right
  ]^{-\scriptscriptstyle{1\over 2}} dP $ and
\begin{eqnarray*}
\delta Y/\delta \zeta = Z^2{d\over dP_m} \int^{P_m}_0 \\
\left \{
\begin{array}{lll}
{\Phi(P) \over 2\pi L^2 \left [1-(P/P_m)\right
  ]^{{\scriptscriptstyle{1\over 2}}}} 
& \int^F_P & 
{EdZ\left (\widetilde{P}\right )d\widetilde{P}
\over 
\left [1-(P/\widetilde{P}) \right ]^{\scriptscriptstyle{1\over 2}}}
\end{array}
\right \}
dP \ .
\end{eqnarray*}
In these terms we must put $Z = Z_1$ the solution given in
(\ref{eq55}).  Thus we can in principle obtain by perturbation theory
a solution to the variable twist problem.



\bsp 

\label{lastpage}



\end{document}